\documentclass[12pt,letterpaper]{article}
\usepackage{epsfig,rotating,setspace,latexsym,amsmath,epsf,amssymb,amsfonts,bm,theorem,cite,caption,subcaption,enumerate,longtable,accents,bbm,appendix,algorithm,algorithmic,graphicx,epsf,authblk,epstopdf,url,color,multirow,mathtools,comment,tabularx,comment,dsfont}
\usepackage[inline]{enumitem}

\newtheorem{theorem}{Theorem}

\newtheorem{lemma}{Lemma}

\newenvironment{Proof}[1]{\medskip\par\noindent{\bf Proof:\,}\,#1}{{\mbox{\,$\blacksquare$}\par}}

\newcolumntype{Y}{>{\centering\arraybackslash}X}
\newcommand{\figref}[1]{\figurename~\ref{#1}}

\allowdisplaybreaks

\title{An Incentive-Compatible Semi-Parallel Proof-of-Work Protocol}

\author{Mustafa Doger \qquad Sennur Ulukus\\
\normalsize Department of Electrical and Computer Engineering\\
\normalsize University of Maryland, College Park, MD 20742\\
\normalsize  \emph{doger@umd.edu} \qquad \emph{ulukus@umd.edu}}

\date{}

\begin{document}

\maketitle

\vspace*{-1.0cm}

\begin{abstract}
Parallel Proof-of-Work (PoW) protocols have been suggested in the literature to improve the safety guarantees, transaction throughput and confirmation latencies of Nakamoto consensus. In this work, we first consider the existing parallel PoW protocols and develop hard-coded incentive attack structures. Our theoretical results and simulations show that the existing parallel PoW protocols are more vulnerable to incentive attacks than the Nakamoto consensus, e.g., attacks have smaller profitability threshold and they result in higher relative rewards. Next, we introduce a voting-based semi-parallel PoW protocol that outperforms both Nakamoto consensus and the existing parallel PoW protocols from most practical perspectives such as communication overheads, throughput, transaction conflicts, incentive compatibility of the protocol as well as a fair distribution of transaction fees among the voters and the leaders. We use state-of-the-art analysis to evaluate the consistency of the protocol and consider Markov decision process (MDP) models to substantiate our claims about the resilience of our protocol against incentive attacks. 
\end{abstract}

\section{Introduction}
Nakamoto consensus \cite{btc-whitepaper}, a solution to the decentralized consensus problem, has a vulnerability against withholding attacks such as double-spending \cite{rosenfeld2014analysis} and selfish mining \cite{selfish-mining} due to the fact that the arrival of the blocks in the longest chain protocol are subject to Proof-of-Work (PoW) scheme which has an exponential distribution \cite{bobtail}. Since an exponential distribution with mean $\frac{1}{\lambda}$ has variance $\frac{1}{\lambda^2}$, adversaries can take advantage of the high variance of the block interarrival times by deviating from the honest behavior. More specifically, whenever it takes longer than usual for honest miners to mine a new block, the adversary can make profits from this by creating blocks and keeping them private until some predetermined conditions on the longest public chain are met \cite{rosenfeld2014analysis, selfish-mining, optimal-selfish}. 

The problem of high variance is not only pertinent to the withholding attacks. Mainstream cryptosystems which rely on Nakamoto consensus with high block interarrival times such as Bitcoin, which produces a block per $10$ minutes, distribute the coinbase reward to the miner who finds a proper nonce. The other miners who could not find a proper nonce are not rewarded and have to start mining the next block to get a chance for the next reward. Although the reward a miner gets is proportional to its hashrate on average, the total number of blocks a miner has to work on until he/she gets a reward is distributed geometrically with a parameter $p=\frac{\text{miners hashrate}}{\text{total hashrate}}$ and has both a high mean $\frac{1}{p}$ and a high variance $\frac{1-p}{p^2}$ for a miner who controls a small fraction of the hashpower in the system. Hence, it is no surprise that a home miner might have to spend decades before getting a reward in Bitcoin system \cite{btc-long-time-unfair}. The presence of network delays and consequent forks exacerbate the problem of unfair reward distribution since blocks belonging to the miners with small fractions of hashpower are less likely to be included in the chain in the long run if they are in a fork race with other blocks \cite{Impact_of_Temporary_Fork, Impact_of_Network_Connectivity_on_Consensus, sakurai2024modelbasedanalysisminingfairness}. As a result, miners who control small fractions of hashpower join their forces together and create pools, which distribute the block rewards among their members regularly, in order to provide a stable income. However, formation of pools is not only a direct threat on the decentralization but also creates vulnerabilities such as different forms of block withholding attacks \cite{courtois2014subversiveminerstrategiesblock, miners_dilemma, fork_after_witholding_attack, power_adjusting}. Moreover, as a pool size grows, the system becomes more vulnerable against selfish mining attacks from that pool. Since the throughput in Nakamoto consensus grows with the arrival of the blocks, it is also subject to limitations that originate from the same core principles as the above such as high variance. 

Many protocols have been suggested to solve some of the above problems of Nakamoto consensus. By modifying how transactions are appended to the ledger, Fruitchain \cite{FruitChains} aims to increase the throughput level and achieve a more fair distribution of transaction rewards among miners with less variance. Sharding techniques such as Prism \cite{Prism} try to increase throughput and reduce scalability and maintenance costs of miners. Bitcoin-NG \cite{bitcoin_ng} increases the throughput and reduces the variance of appending transactions to the ledger with a PoW leader election mechanism. GHOST rule \cite{Sompolinsky2015SecureHT} allows miners of the orphan blocks to get partial rewards by replacing the longest chain rule and mildens the unfair reward distribution. Modifications to fork choice rule of Nakamoto consensus such as \cite{selfish-mining, Publish_or_Perish, decor+, preneel_common_metrics} try to mitigate the incentive attacks on reward mechanism. PHANTOM \cite{PHANTOM_GHOSTDAG} and SPECTRE \cite{SPECTRE} protocols allow blocks to have multiple parents creating Directed Acyclic Graphs (DAG). As a result, higher block generation rates, $\lambda$, are allowed which in turn creates a more fair blockchain environment with high throughput. Multistage PoW \cite{multistage_pow} divides the PoW problem for each block into multiple sequential stages which decreases the variance of block interarrival times. Many of these protocols also come with limitations such as low fault tolerance, absence of rigorous security violation analysis, absence of an efficient, global and consistent ordering of transactions, communication overheads and resilience against possible withholding attacks \cite{PHANTOM_GHOSTDAG, SPECTRE, bitcoin_ng_greedy_mine, bitcoin_ng_incentive_revisited, multistage_pow_props_vulnerabls, Spy_Based_multistage_pow}.

Parallel PoW \cite{parallel-pow-bounds} has a similar principle where the system as a whole needs to solve $L$ puzzles in parallel, which decreases the variance of reward distribution times and increases resilience against double-spending by decreasing the variance of block interarrival times. However, in its raw form, parallel PoW protocols are more vulnerable to incentive attacks \cite{tailstorm, dag_parallel_pow}. Bobtail \cite{bobtail} can be seen as a modification to the raw form of parallel PoW scheme to increase the resilience against proof withholding attacks. Tailstorm and DAG-style voting schemes are other modifications introduced in \cite{tailstorm, dag_parallel_pow} that change the reward distribution rule between the parallel proofs to punish withholding attacks in the raw form of the parallel PoW scheme. It is worth noting that, parallel PoW protocols where miners create parallel proofs on top of the same block are distinct from sharding techniques where miners create and maintain parallel chains. In this paper, we only consider parallel PoW protocols.

We first consider the existing variations of the parallel PoW protocols \cite{bobtail, tailstorm, dag_parallel_pow} and develop hard-coded incentive attacks that show shortcomings of the protocols. For Bobtail \cite{bobtail}, drawing an analogy to the analysis of Zhang and Preneel \cite{preneel_common_metrics} about smallest hash based fork choice rules, we consider two types of attacks and argue that the profitability threshold, $\alpha_T$, i.e., the minimum fraction of hashpower needed for a profitable attack, is zero. For Tailstorm \cite{tailstorm}, our hard-coded proof withholding attack results in a profitability threshold of $\alpha_T\leq\frac{1}{L}$ and results in revenues as large as the upper bound of the profit of a selfish miner in Nakamoto consensus. For DAG-style voting \cite{dag_parallel_pow},  even when the adversary has no network influence, i.e., $\gamma=0$, contrary to the authors' intention, when adversarial fraction of hashpower $\alpha>\frac{1}{3}$, the punishment mechanism favors the adversary for a hard-coded attack we develop and the resilience is worse than the resilience in Tailstorm in this parameter regime. For all three schemes, we verify our theoretical analysis and arguments with simulations.

Next, we introduce a voting-based semi-parallel PoW protocol that outperforms the existing models from most practical perspectives. Existing works on parallel PoW not only lack resilience against incentive attacks but also have communication overhead problems. Moreover, arbitrary increase of throughput without conflicts among transactions is not possible. Our protocol offers a solution for all these problems where each PoW solution (proof) is a vote and proofs can refer to each other as in \cite{bobtail, dag_parallel_pow, tailstorm}. Additionally, each vote points to a proposal ledger through its Merkle root, to be voted in the next block. The communication overhead is decreased by requiring the vote to keep the proposal ledger hidden but specify the amount of transaction fees offered. 

We introduce several rules for honest miners to make sure all miners of size $\alpha<0.5$ pick the proposal ledger that offers the most fees as $L$ increases. As a result, only one hidden ledger is shared and communication overheads as well as transaction conflicts are avoided. We introduce punishment mechanisms as in \cite{preneel_common_metrics, Publish_or_Perish, decor+} to further reinforce incentive compatibility of the protocol as well as a fair distribution of transaction fees among the voters and the leaders. The introduction of ledger choice rule and split rule for the distribution of transaction fees makes sure that arbitrary manipulations of ledgers are not feasible for adversaries. We adapt the state-of-the-art security latency analysis of Nakamoto consensus \cite{guo-btc-sec-lat, our-sec-lat-extended} to analyze the double-spending probability for our protocol. Further, we consider Markov decision process (MDP) models to substantiate our claims about the resilience of our protocol against incentive attacks. 

In the rest of this paper, we assume a single colluding adversarial entity which holds $\alpha$ fraction of the hashpower in the system. We assume a network of $n$ honest miners connected to each other with some network topology and large $n$. We denote the network influence of the adversary, i.e., the fraction of honest miners that prefer the adversarial chain in case of a fork as $\gamma$. We abstract the mining puzzles with a collision-free hash function, $f_H$, which acts as a random oracle. In Section \ref{sec::prev_parallel_pow}, we summarize the existing variations of the parallel PoW protocols and show their vulnerabilities against incentive attacks. In Section \ref{sec::our_protocol}, we formally introduce our protocol and subsequently analyze it in Section \ref{sec::analysis_of_our_protocol}. In Section \ref{sec::discussion}, we discuss how our protocol relates to the existing work on PoW based protocols in general, its limitations and possible future work.

\section{Incentive Attacks on Existing Variations of Parallel PoW Protocols} \label{sec::prev_parallel_pow}
In this section, we consider proof withholding attacks on the existing variations of parallel PoW protocols, where we pick the relative rewards as the relevant metric which is the fraction of the rewards an agent has in the system as commonly used in most of the literature. These attacks result in relative rewards higher than the honest mining for many paramater regimes and show that existing variations of parallel PoW protocols are even less resilient against incentive attacks than the Nakamoto consensus in many realistic settings. 

We start by defining the raw form of the parallel PoW protocol. In the raw form of the parallel PoW protocol, a block consists of $L$ PoW solutions (simply referred as proof or vote) that build upon the previous block by pointing to its hash. Each proof contains semantic information such as the previous hash, Merkle root of a ledger, the reward address etc., which we ignore. From a practical point of view, we only care for the hash value of the proof and the pointer to the previous hash. In the raw form, the proofs refer to the previous block and the hash value of these proofs have to be below a certain threshold. All PoW votes that confirm the same parent block are eligible to be included in the next block and $L$ of them are needed to create the next block. As a result, given the most recent block, the proofs of the next block can be generated in parallel. This in turn allows a full cooperation between miners without the need for a pool. 

In the honest protocol, essentially everyone works on top of the most recent block and miners immediately release the proof they mine, which is essentially a vote (or a validation) on the previous block. In case of a fork, the miners pick the chain with the most aggregate vote to build upon. Thus, the first time $L$ new proofs are mined on top of a previous block, a new block can be created and revenues can be shared between $L$ proofs. However, this usually requires an election rule to decide who will combine the proofs to create the new block. A simple way to do that is to let the miner of the proof with the lowest hash value be the leader and create the next block. For the purpose of the analysis we perform in this section, we believe that the details provided here are sufficient to build upon. For further details on the raw form of parallel PoW, we refer the reader to \cite{parallel-pow-bounds}.

If the adversary does not release its proofs for the current block immediately as they are mined, while being able to observe the honest proofs immediately as they are mined, this informational advantage can potentially result in extra profits. This type of attacks are called proof withholding attacks. Next, we give an overview of how existing parallel PoW modifications \cite{bobtail, tailstorm, dag_parallel_pow} work and develop hard-coded incentive attacks that result in more gain than the honest mining.

\subsection{Bobtail}
In Bobtail, each proof has a hash value and can be created in parallel as usual and any proof with a hash value lower than a certain threshold is a valid proof.\footnote{In the actual paper of \cite{bobtail}, the average hash value of $L$ proofs needs to be below a threshold, however, for practical purposes and simplicity, we ignore proofs above a certain threshold and assume any proof below it is a valid proof as in \cite{parallel_pow_red_variance}.} The proof that has the lowest hash value is allowed to combine the current $L$ proofs to create a block. To prevent the proof withholding attacks, each proof has a support value, which is the main difference from the raw form. This \emph{support value} is essentially a pointer to the lowest hash valued proof it has seen so far among the proofs for the current block height. To create a block, a lowest hash proof is only allowed to combine the proofs that have higher hash and support hash value. 

We note that the authors of \cite{parallel-pow-bounds} state that Bobtail is not a parallel PoW protocol. However, as both protocols allow parallel mining of the proofs and the essential difference between the raw form of the parallel PoW protocol and Bobtail is the support value that proofs in Bobtail contain, in this paper, we treat it as a variation of parallel PoW protocols.

\begin{figure}[t]
    \centerline{\includegraphics[width=0.7\columnwidth]{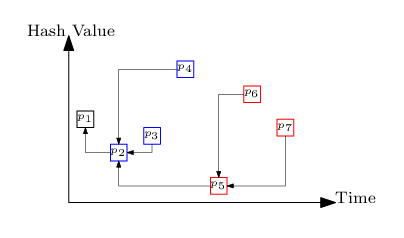}}
    \caption{Bobtail protocol proof model.}
    \label{fig::bobtail}
\end{figure}

In \figref{fig::bobtail}, we show an example of how proofs point to each other according to their mining times and hash values. Each proof, denoted by $p_i$, has a hash value aligned with $y$ axis and a mining time aligned with $x$-axis and a pointer to its support hash. For example, since $p_5$ has the smallest hash value among all, if $L=7$, then the miner of $p_5$ is allowed to create a block with references to all $p_i$. On the other hand, the miner of $p_2$ cannot use any of the red proofs to create a block since they have smaller hash values of their own or smaller support value.

Without the support value rule, the adversary can mine but hide all its proofs until it can form a proof set of size $L$ containing all of its own proofs and all honest proofs that have higher hash value than its own proofs. The attacker cannot use the honest proofs that have lower hash value than its own lowest hash value while creating a block. With the support value rule in place, this proof withholding attack is mitigated as the attacker cannot use the honest proofs that have lower hash value than its own lowest hash value while creating a block, but all the proofs that refer to them are not eligible either. For example in \figref{fig::bobtail}, if $p_2$ is mined by the adversary and it tries to create a block with $p_2$ having the lowest hash, it cannot use $p_5$, $p_6$ or $p_7$ when creating a block even though $p_6$ and $p_7$ have higher hash value than $p_2$.

To further strengthen the protocol, the authors consider a bonus reward rule where proofs, whose support proof is the lowest hash valued proof which eventually created the final block, get a bonus reward. The protocol also defines further incentive rules about prioritizing inclusion of proofs in case multiple sets of $L$ proofs are possible; see \cite{bobtail} for further details.

In Appendix~\ref{app::bobtail},  we first provide details of the two attacks considered in the original paper of Bobtail, and later create a stronger hard-coded attack for which a pseudo-code is also given. We also provide both a theoretical overview as well as a simulation result which show Bobtail simply has profitability threshold of $\alpha_T=0$ due to the informational advantage that the adversary possesses. As a result, even a small home miner can mount an incentive attack on the protocol. It is also noted that the attack scales block interval times,  decreases the throughput and increases the latency.

\subsection{Tree- and DAG-Style Voting}
Tailstorm \cite{tailstorm} is a tree-style parallel PoW protocol \cite{dag_parallel_pow}. Tailstorm introduces two main innovations compared to the raw form of the parallel PoW protocol. First, the proofs that are being generated for the current block height can refer to each other, creating a tree of proofs. As long as the root of the tree is the previous block, they are all eligible to be combined together to create the next block. Tailstorm further deviates from the raw form in how it rewards the proofs. Each block is rewarded proportional to the maximal depth of its proof tree and the reward is distributed equally to each proof in the tree. In Tailstorm, as every proof in a block gets rewarded equally, this also means that they are punished equally, since they can get more rewards if they build their proofs in a sequential chain and if a proof creates a fork, the reward for every proof inside the block decreases equally.   

The addition of this tree  structure and the new reward mechanism aim to mitigate the proof withholding attacks. DAG-style voting \cite{dag_parallel_pow} introduces modifications along the same lines that aim to fix the unfair punishment mechanism in Tailstorm. In DAG-style voting, proofs are allowed to refer to multiple proofs which results in DAG. In this case, the reward of each proof is proportional to the total number of ancestor and descendant proofs it has within the block, which punishes the non-sequentiality more fairly; see \cite{dag_parallel_pow} for further details.

Both the authors of \cite{tailstorm} and \cite{dag_parallel_pow} consider proof withholding attacks with a normalized reward metric which is the reward of an attacker up to the current block divided by the progress of the chain. Similar metrics are also used in recent works such as \cite{intermittent_mining,profit_lag}. They show the resilience of their algorithm according to this normalized reward metric. However, in the long run, what really matters is the relative metric that directly represents the market share an agent holds as originally analyzed for Nakamoto consensus \cite{intermittent_mining,grunspan2019-profitability-selfish-mining,profit_lag,Grunspan_witholding_resilience}. In fact, the analysis of \cite{tailstorm,dag_parallel_pow} ignores the zero-sum game principle \cite{selfish-mining,mind_the_mining} used to analyze mining strategies in systems consisting of multiple agents. In Appendix~\ref{app::relative_reward}, we explain in detail why the relative reward metric is the better choice when analyzing incentives in PoW protocols. 

In Appendix~\ref{app::witholding_attack}, we handcraft proof withholding attacks to show vulnerabilities of tree- and DAG-style voting protocols, where we consider the relative reward metric. The attacks we consider are quite similar for both tree- and DAG-style voting protocols, where the attacker simply withholds its own proofs which results in an informational advantage. This informational advantage simply results in a direct revenue advantage in Tailstorm, whereas the advantage in DAG-style voting protocol requires the adversarial power to be above some threshold. We explain the attack, analyze it theoretically and then give a pseudo-code for a practical implementation. To further strengthen our analysis, we simulate the attack to verify our theoretical results.

With the hard-coded proof withholding attack, the fraction of the adversarial proofs in the public chain reaches $\frac{\alpha}{1-\alpha}\frac{L-1}{L}$ even when $\gamma=0$, for both tree- and DAG-style voting. Since this fraction directly equals the relative reward ratio in Tailstorm, an upper bound on the profitability threshold, i.e., an upper bound on the minimum adversarial power needed for a profitable attack is $\alpha_{T}\leq\frac{1}{L}$. In other words, if the adversary holds more than $\frac{1}{L}$, then the attack is always profitable. For less adversarial power, better attack strategies may still exist. Although parallel PoW protocols become more secure against double-spending attacks as $L$ increases and the income of the miners become more stable since even the small home miners \cite{btc-long-time-unfair} can get a reward as the nonce difficulties are easier, the above attack implies, Tailstorm becomes more vulnerable against the incentive attacks as $L$ increases. For DAG-style voting, the proof withholding attack is still in place, however, the profitability threshold is different since proofs get rewards proportional to the number of ancestors and descendants. 

\subsection{Numerical Results for the Incentive Attacks}

\begin{figure}[t]
	\centerline{\includegraphics[width=0.7\columnwidth]{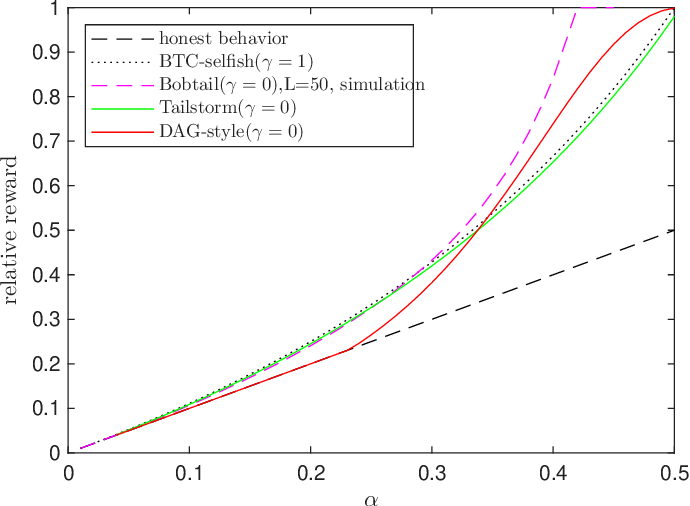}}
	\caption{Relative revenues of withholding attacks on existing variations of parallel PoW protocols.}
	\label{fig::withhold_attacks}
\end{figure}

We plot the relative reward ratio of the attacks for the simulation of Bobtail and the theoretical results of Tailstorm \cite{tailstorm} and DAG-style protocol \cite{dag_parallel_pow} in \figref{fig::withhold_attacks}, where we pick $L=50$ and $\gamma=0$. For reference, we include the honest mining curve, i.e., when the adversary mines honestly and gets $\alpha$ rewards and the upper bound of selfish mining in Bitcoin, or more generally in Nakamoto consensus, where $\gamma=1$, which is equal to $\frac{\alpha}{1-\alpha}$. We truncate the curves before they rise above the honest mining strategy for the sake of neatness.

Bobtail has a profitability threshold of $\alpha_T=0$, even though it is hard to discern the curves for small $\alpha$ in the figure. When $\alpha\geq \frac{1}{3}$, Bobtail gives even more profits than the upper bound of the selfish mining in Nakamoto consensus. We note that, we added a prevention mechanism to our simulation of Bobtail that makes sure honest miners do not wait too long to find a small hash value as explained in Appendix~\ref{app::bobtail}. Despite this, the hard-coded attack is stronger than the other attacks in the other protocols.

It is clear that for $\alpha>\frac{1}{L}$, the attack on Tailstorm is more profitable than honest mining as everyone including the honest miners is punished equally and the limit as $\alpha\to0.5$ is $\frac{L-1}{L}$. The behavior of the curve is similar to the upper bound of the selfish mining curve of Nakamoto consensus when $\gamma=1$.

For DAG-style voting, the introduction of reward scaling according to the depth of the graph makes the protocol more resilient against incentive attacks than the Tailstorm only for small $\alpha$ and the profitability threshold in DAG-style voting is larger than Tailstorm. However, even when $\gamma=0$, the DAG version of the attack becomes more profitable than the same withholding attack in the tree-style voting for $\alpha>\frac{1}{3}$. This is due to the fact that the rewards are scaled according to the ancestor and descendant counts and the adversarial proofs have more than the honest ones have when $\frac{\alpha}{1-\alpha}>0.5$. This is contrary to the intention of the authors as they introduced reward scaling to punish adversarial actions since they hold less than $0.5$ fractional power. 

To see this, we could also consider a simple logic where honest proofs only point to honest proofs and adversarial proofs only to adversarial proofs. In this case, the adversary would get more rewards if it has more proofs, which happens when $\alpha>\frac{1}{3}$, since around $\alpha$ fraction of proofs which are honest are replaced by adversarial proofs, hence the adversary has $\alpha$ fraction of $1-\alpha$ proofs. This also explains why the profit of this attack for $\alpha>\frac{1}{3}$ (even when $\gamma=0$), is more than $\frac{\alpha}{1-\alpha}$, which is the upper bound of the profit of selfish mining attack in Nakamoto consensus \cite{optimal-selfish}, since rewards are proportional to the square of the number of proofs honest and adversarial agents hold. It is worth noting that we did not perform an exhaustive search on all possible attacks and provided only a hard-coded attack. The actual profitability threshold might be even lower.

For example, in \figref{fig::dag_2_attack} of Appendix~\ref{app::witholding_attack}, when the DAG attack starts, initially at block $h$ and $h+1$, the reward scaling is more beneficial to honest miners. However, as the adversary gains momentum in next blocks and more honest proofs are replaced in a block, honest miners get punished more. In fact, we could consider an enhanced DAG withholding attack which has even more profits than the original attack in DAG for $\alpha>\frac{1}{3}$, where the adversary avoids pointing to honest proofs whenever it can. Since our goal in this section is to show the vulnerability of the existing parallel PoW protocols and its inferior resilience compared to the resilience of Nakamoto consensus, the results presented here and the analysis in Appendix~\ref{app::bobtail} and Appendix~\ref{app::witholding_attack} are sufficient. 

The handcrafted attacks we consider work irrespective of the $\gamma$ value, hence better attacks exist that take advantage of $\gamma$. Moreover, the handcrafted attacks are non-dynamic in the sense that the adversary does not change its strategy even when it encounters a high variance mining time. For example, in selfish mining attack of Nakamoto consensus, the profit of the attack comes from the fact that the adversary takes advantage of the high variance of the mining times, i.e., whenever the honest miners encounter a high block interarrival time, the adversary has an opportunity to be conversely lucky \cite{bobtail}. For example, since rewards are scaled according to the depth, the adversary could benefit from dynamically switching to honest mining for blocks for which it has created less than usual proofs. 

\section{Proposed Protocol Description}\label{sec::our_protocol}
We start by questioning the extent of parallelness needed in a parallel PoW protocol as well as the efficiency of creating transaction ledgers associated with the proofs. If we think of each proof as a small block with some transactions attached to it as in \cite{tailstorm,dag_parallel_pow}, the propagation times associated with the ledgers of proofs would imply that more proofs are created during the network delays. In other words, the system has more parallel proofs. Hence, such a protocol needs a high level of parallelness, i.e., proofs that are mined at around the same time should be valid and compatible to be combined together. However, this also implies that the protocol usually would have throughput loss due to the conflicts in ledgers of the proofs mined in parallel. In essence, this approach is more or less similar to the approach of GHOST and PHANTOM protocols \cite{Sompolinsky2015SecureHT,PHANTOM_GHOSTDAG}, where authors consider block trees and DAGs to make the blocks mined in parallel compatible with each other where nonce difficulties can be decreased to increase the throughput. However, we have seen that within the framework of parallel PoW protocols tree- and DAG-style votings do not have resistance against withholding attacks. Further, the withholding attacks end up changing the confirmed set of transactions since the set of confirmed proofs change which have different transaction ledgers in \cite{dag_parallel_pow,tailstorm}. On the other hand, a protocol like Bobtail \cite{bobtail} separates the proofs form the ledgers and offers a leader election mechanism. This approach decreases communication overhead and solves the problem of conflicts between ledgers. However, it is well-known that smallest hash based mechanisms used in Bobtail allow attackers to have informational advantage which in turn can be converted into profits with withholding attacks. Further, smallest hash proof withholding attacks also increase block interarrival times.

Notice that, if a protocol separates the proofs from the ledgers, it decreases propagation times of proofs. As a result, less proofs will be mined during delay periods and thus the level of parallelness does not need to be high. This however, requires a ledger election mechanism that is incentive compatible. Further, in this case, ledger conflicts can be omitted with a clever mechanism. In this section, we explain a protocol that allows a reasonable amount of parallelness with small proof propagation times, an incentive compatible leader election mechanism, less communication overhead and has no ledger conflicts. Further, it has superior variance properties compared to Nakamoto consensus, i.e., a more fair distribution of coins, better safety guarantees and resilience against withholding attacks and throughput increase. 

\subsection{Blocks and Chain Structure} 
\begin{figure*}[t]
    \centerline{\includegraphics[width=1\columnwidth]{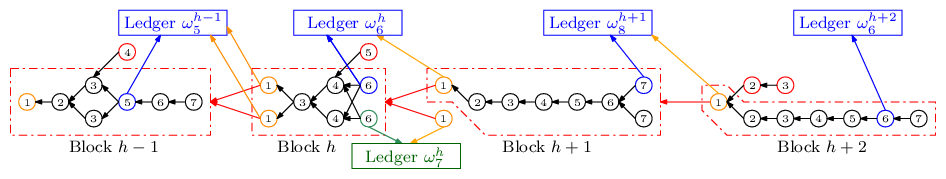}}
    \caption{The proof model of the semi-parallel PoW protocol with $L=7$. Orange and black circles represent the initiator and incremental proofs and the numbers inside the circles represent the incremental heights. Each initiator proof refers to a ledger from the previous block height as the winning ledger. Blue boxes represent the winning ledgers based on the most aggregate work fork choice rule. Here, if the proof $7$ of block $h+2$ is the most recent proof with most aggregate work and $b=1$, the ledgers until block height $h+1$ are confirmed, since they are part of the most aggregate work chain. But, ledger $\omega_6^{h+2}$ is not $b=1$-confirmed since $1$-confirmation requires $1$-complete block on height $h+3$ where the initiator of block $h+3$ needs to refer to $\omega_6^{h+2}$ as the winning ledger. Red dotted boxes (e.g., block $h$) together with a valid ledger (e.g., $\omega_6^h$) are considered as a complete block. Note that, block $h$ with ledger $\omega_7^h$ is a complete but orphan block. To replace the block $h$ and its associated ledger $\omega_6^h$ with the block $h$ and its associated ledger $\omega_7^h$ in the view of honest miners, the adversary creates an initiator proof that refers to $\omega_7^h$. 
    The adversary has to create further proofs and share complete blocks at block height $h+1$ and $h+2$ to make the block $h$ and its associated ledger $\omega_7^h$ as the fork choice rule (most aggregate work). Note that the adversary cannot use any proofs of block $h+1$ or $h+2$ in the figure since initiator proof of the block $h+1$ refers to $\omega_6^h$ as the winning ledger and all the subsequent proofs build on top of this initiator proof. Thus, to rewrite the history, adversary has to work alone once it falls behind in the race of most aggregate work.}
    \label{fig::our_parallel_protocol}
\end{figure*}

We consider a series of blocks chained together as depicted in \figref{fig::our_parallel_protocol}. A valid block at height $h$, represented with red boxes in the figure, consists of $L$ or $L+1$ valid PoW solutions (proofs) clustered together, i.e., $\omega^{h}_i$, $i=1,\ldots, (L~ or~L+1)$. Proofs are divided into two categories: initiators and incrementals. The orange circles represent the initiators whereas other circles are incrementals in the figure. 

A valid initiator proof consists of 6 parts concatenated together, i.e., $\omega^h_1=(\omega^h_{1,1},\ldots,\omega^h_{1,6})$ where w.l.o.g. we assume the miner who mines $\omega^h_1$ is the miner $i=1$.  The sixth part of the proof, i.e., $\omega^h_{1,6}$, contains the nonce to make sure $f_H(\omega^h_1)<T_L$. The fifth part, i.e., $\omega^h_{1,5}$, is simply the hash of the coin address of the miner $i=1$ to store its mining rewards in case its solution is accepted. The fourth part, $\omega^h_{1,4}$, is the Merkle root of the transaction ledger proposed by miner $1$, which will be explained in detail later. The third part, $\omega^h_{1,3}$, is simply the total accumulated transaction fees of the proposed ledger associated with $\omega^h_{1,4}$. The second part is a summary of (w.l.o.g. assume $L$) clustered proofs of preceding block height, i.e., $\omega^h_{1,2}=\omega^{h-1}=(f_H(\omega^{h-1}_{1}),\ldots,f_H(\omega^{h-1}_{L}))$. If two initiator proofs at height $h-1$ exist, say miner $1$ and miner $2$ created valid initiator proofs $\omega^h_1$ and $\omega^h_2$, then, they need to agree about the history to be clustered together in a block in order to prevent double-spending, i.e., $\omega^{h-1}_{1,2}=\omega^{h-1}_{2,2}$ ($\omega^{h-1}_{1,1}=\omega^{h-1}_{2,1}$ should also be satisfied, which will be explained in Section~\ref{sec::ledger-election}).

A valid incremental proof created by miner $i$ consists of 6 parts concatenated together, i.e., $\omega^h_i=(\eta^h_{i,1},\ldots,\eta^h_{i,6})$. Similar to the initiator proofs, the last four parts of the incremental proofs, i.e., $\eta^h_{i,3}$ to $\eta^h_{i,6}$ have the same rules and serve the same purpose. Thus, in this paper, we simply refer to these parts with $\omega$ instead of $\eta$ for the sake of simplicity. However, the first part of the proof is simply a number representing the incremental work within the cluster of proofs at block height $h$, hence we call it the incremental height of the proof. The second part of the proof, related to that, is a list of summary of proofs of the previous incremental height, with the condition that the list can contain at most two proofs and these proofs agree about the history, be it the ledger or the incremental proofs. To simplify the notation, we refer to the miner of the proof at incremental height $i$ as miner $i$. An initiator has incremental height $1$ by default and all incrementals inherit $\omega^h_{1,1}$ and $\omega^h_{1,2}$ of the initiators within the cluster. 

\subsubsection{Example Proof Model}
As an example, consider the block at height $h$ in \figref{fig::our_parallel_protocol} with $L=7$. Two valid initiator proofs are represented as orange circles created by miner $1$ and $2$, with $\omega^{h}_{1,2}=\omega^{h}_{2,2}$ and $\omega^{h}_{1,1}=\omega^{h}_{2,1}$, which were possibly created approximately at the same time, hence were not aware of each other, where we order them with respect to their summary, i.e., $f_H(\omega^{h}_{1})<f_H(\omega^{h}_{2})$. 

In this figure, the black circles represent the incremental proofs and the numbers inside the circles represent the incremental heights. The next incremental proof, let us call its creator miner $3$ -- who is honest and was aware of both of the initiator proofs at the time of the mining, will have $\eta^h_{3,1}=3$, $\eta^h_{3,2}=(f_H(\omega^{h}_{1}),f_H(\omega^{h}_{2}))$. Here, we note that although the proof height of $\omega_3^h$, i.e., the height of the proof within the block height $h$, is $2$, the incremental height is $3$ since the incremental work is $3$. 

Next, assume two more incremental proofs are created by miner $4$, $5$, such that $\eta^h_{i,1}=4$, $\eta^h_{i,2}=f_H(\omega^{h}_{3})$ for $i=4,5$. Since these two proofs agree about the history, the next miner who is aware of both, call miner $6$, can combine them as well, i.e., $\eta^h_{6,1}=6$, $\eta^h_{6,2}=(f_H(\omega^{h}_{4}),f_H(\omega^{h}_{5}))$ where they are ordered according to $f_H(\omega^{h}_{4})<f_H(\omega^{h}_{5})$. In a similar way, any other proof $i$, say $i=7$, which has $\eta^h_{i,1}=6$, $\eta^h_{i,2}=(f_H(\omega^{h}_{4}),f_H(\omega^{h}_{5}))$ can be combined with the proof of miner $6$ on the next incremental height. On the other hand, a miner $5'$, possibly an adversary or an honest miner who was aware of the proof $\omega^{h}_{4}$ but not aware of the proof $\omega^{h}_{5}$, may have created the proof $\omega^{h}_{5'}$ with $\eta^h_{5',1}=5$, $\eta^h_{5',2}=f_H(\omega^{h}_{4})$ represented as red circle. In this case, we do not allow the proof of the miner $5'$ to be combined with the proof of the miner $6$ and the next honest miner will prefer combining the proofs of $6$ and $7$ since it has more aggregate work. Thus, the proof of the miner $5'$ becomes orphan and is not included in $\omega_{1,2}^{h+1}$, represented as the red dash dotted box at block $h$. 

In a similar manner, at block height $h-1$, the proof represented with a red circle cannot be combined with the proof represented with a blue circle, since their parent sets differ. Regarding the red circles of block height $h+2$, multiple scenarios are possible. For example, an adversary may have kept both private until two parallel honest proofs are generated. In this scenario, the red and black proofs with incremental height $3$ cannot be combined  since they have different parents and the fork resolves later on according to the fork choice rule described in Section \ref{sec::fork_choice_rules}.

To summarize this rule, the two incremental proofs at incremental height $i$ who have the same parents, can be combined in the next proof height resulting in incremental height $i+2$. This also explains why $L+1$ PoW solutions for a block as in block $h+1$ in \figref{fig::our_parallel_protocol} is allowed as the last incremental height can have two proofs which can be combined by the initiator of next block height. We call two proofs that can be combined as \textit{parallel proofs}.

\subsection{Ledger Election}\label{sec::ledger-election}
It remains to define $\omega^h_{i,1}$, i.e., the first part of an initiator PoW solution. Before doing so, we make some observations. It should be clear that for height $h$, each miner has a proposed ledger that does not necessarily agree with others. However, miners of the same cluster agree with each other about the previous block since they either build on top of each other or $\omega^h_{i,2}=\omega^h_{j,2}$ for two initiator proofs of the same cluster. 

Thus, it remains to pick a rule to decide about the ledgers. If we are to accept all the transactions of the proposed ledgers of a clustered proof at height $h-1$, we need to further decide about a rule in case of conflicts, which can be as simple as first ordering by incremental height within the cluster and accepting the ledgers in the order of their hashes (or Merkle roots) in order to ignore further conflicts. However, this will both increase the communication overhead and also create a necessity of a further rule in case a ledger is not shared by the miner, e.g., a ledger withholding attack by the adversary. 

From $\omega^h_{i,4}$, it should be clear that within the cluster, each miner has a proposed ledger, and the ledger pays $\omega^h_{i,3}$ in transaction fees in addition to the coinbase reward. If we only accept one ledger, we need each proof of a cluster at height $h$ to agree with each other about the ledger of the previous block height, in order to prevent rewriting of the history, i.e., double-spending. Thus, we consider the following scenario: The transaction fees associated with $\omega^h_{i,4}$ are offered to be shared with the miners of the proofs for the next block height $h+1$.  If this is the case, it is in the best interest of miners at block height $h+1$ to pick the ledger of the previous height that pays the most fees. 

Thus, the rule is to simply pick the ledger that pays the most among the valid ledgers of a cluster from the previous height, i.e., $\omega^h_{i,1}=\omega^{h-1}_{j^*,4}$ where $j^*=\arg\max_j\{v(\omega^{h-1}_{j,3})\}$.  Here, $v(\omega^{h-1}_{j,3})=0$ if $\omega^{h-1}_{j,4}$ is not valid or withheld, else $v$ is simply the identity function.

If parallel initiator proofs exist in the same cluster at block height $h$, i.e., $\omega^{h}_{1,2}=\omega^{h}_{2,2}$, then we require them to agree about their ledger pick for the previous height as well, i.e., $\omega^{h}_{1,1}=\omega^{h}_{2,1}$. As all incrementals build on top of the initiators, they necessarily inherit the pick of the initiators, hence we have a consensus about the ledger of the previous block height among the proofs clustered together. 

\subsubsection{Example Proof Model} In \figref{fig::our_parallel_protocol}, the proofs that offer the most fees are represented with blue circles which are pointed by their creator as well as the initiator proof of the next block height. Furthermore, notice that, both initiator proofs at block $h$ pick the ledger of the proof $\omega_5^{h-1}$, hence they can be combined. On the other hand, even though the initiator proofs at block $h+1$ agree about the proofs in the previous block $h$, they pick different ledgers from the previous block, thus they cannot be combined. Further, any incremental proof and blocks mined on top of one of these incremental proofs cannot be used by the other incremental proof since they inherit the ledger choice of the incremental proof. Thus, to rewrite the history, the adversary has to create a separate proof tree with more aggregate work.

\subsection{Communication Overhead}
As a valid PoW solution contains the amount of the fees the associated proposal ledger offers, i.e., $\max_j\{\omega^{h-1}_{j,3}\}$, there is an open competition about the leader (ledger) of a cluster, i.e., the miner whose proposed transaction ledger is going to be accepted. Thus, the leader of a cluster at block height $h-1$ is essentially decided by the winning cluster of the next block height $h$. Miners of a cluster hide their proposed ledger until there are $L$ proofs (votes) accumulated in the cluster in order to prevent others from replicating their proposal (if it pays a high fee). This in turn decreases the communication overhead associated with parallel PoW protocols, since honest miners only need to share their ledger if they pay the most fee and only when there are $L$ votes accumulated.

As an incremental proof contains on average $(6+\mathbb{E}[\Bar{s}])\times 32$-bytes, where $s=1+\Bar{s}$ refers to the number of combined incrementals of the previous proof height with $1 \leq s\leq2$, we can assume that it will be received fast and will not congest the network. Although  initiator proofs contain $(L+\mathbb{E}[\Bar{r}]+5)\times 32$-bytes, where $r=1+\Bar{r}$ refers to the number of combined proofs of the previous block height with $1 \leq r\leq2$, they are much less frequent to have a considerable impact on the network congestion.

\subsection{Throughput Increase}
We assume that the ledger size is limited, however, to increase throughput, the maximum allowed size of the ledger associated with a proof of a miner $i$ is proportional to the height of the proof within the cluster, which we call proof height. More specifically, the proposed ledger size of the PoW solution $\omega^{h}_i$ is limited by $\lceil\frac{C_h(\omega^{h}_i)}{M}\rceil \times B$-byte, where $C(\omega^{h}_i)$ is the proof height. Note that, $C(\omega^{h}_i)$ is different than $\eta^h_{i,1}$ as $\eta^h_{i,1}$ is the incremental height. 

For example, if an incremental block has two initiator blocks as its parent, then it has incremental height $\eta^h_{i,1}=3$ but  $C(\omega^{h}_i)=2$. Thus, there are at least $\frac{L}{2M}$ and at most $\frac{L}{M}$ quantized throughput levels. As a result, miners are incentivized to share their proof as soon as they find a valid one to minimize the number of parallel proofs since it allows the cluster as a whole to increase the number of transactions that can be included in a proposal ledger. 

Further, the proposed ledger is created by a single miner which makes sure that no conflict can occur, i.e., unlike the other parallel PoW protocols throughput is not wasted. This rule also reinforces a decreased communication overhead as it is unlikely for a ledger of a proof with small $C(\omega^{h}_i)$ value to offer a high fee, $\omega^{h}_{i,3}$, hence miners simply focus on creating votes initially until high throughput levels are achieved, which is when they start to focus on creating candidate ledgers for the next block height.

\subsection{Ledger Content}
A valid ledger created at block height $h$ needs to satisfy certain conditions related to the mining rewards and fees. As the winning ledger of a cluster at block height $h-1$ and other miners within that cluster are pointed by the miners of height $h$, the coinbase rewards of height $h-1$ are distributed in the ledger of the block height $h$. Hence, a ledger at the block height $h$ needs to contain the coinbase reward distribution of height $h-1$ in order to be considered as valid. 

Further, as the transaction fees of the winning ledger of height $h-2$ are shared between the miner who created that ledger and miners of winning cluster at block height $h-1$, which are pointed by miners of height $h$, those fee payments have to be finalized in the ledger of height $h$ as well, which may require a smart contract algorithm involving the associated ledgers and the miners' addresses of the previous two heights. The fraction of transaction fees that will be paid to the ledger creator is denoted as $r$ and is explained in Section \ref{sec::transaction_fee_split_rule}.

In short, one can think of each proof on block height $h$ as a vote on the proposed ledgers of height $h-1$. After a while as the votes at height $h$ are accumulated enough, the voters also start to focus on proposing ledger candidates for the next block height which come together with the vote. Each vote gets rewarded for their work through the mining rewards and transaction fees.

\subsection{Difficulty Adjustment, Fork Choice and Block Confirmation Rules}
\label{sec::fork_choice_rules}
For difficulty adjustment, we simply assume that $T_L$ is updated every two weeks to make sure that the expected number of blocks generated every two weeks, call it $N$, stays the same. Since each block has on average $L+\mathbb{E}[\Bar{s}]$ PoW solutions, this implies that the difficulty is adjusted such that, on average, $(L+\mathbb{E}[\Bar{s}])\times N$ proofs are generated every two weeks. Since the protocol requires a semi-parallel PoW, where at least $\frac{L}{2}$ proofs are sequential, the variance of block interarrival times decreases with $\mathcal{O}(\frac{1}{L})$ compared to Nakamoto consensus \cite{serial_vs_parallel_pow}.  

At all times, the miners pick the chain that contains the most aggregate work (heaviest chain). The aggregate work of a proof is the total number of proofs, weighted by their respective  nonce difficulties it has in its history, plus the work of a compatible parallel proof if it exists. Here, without loss of generality, we assume that all the weights are the same. For example, the aggregate work of an incremental proof $\omega^{h}_i$ is the total number of proofs until block height $h$ plus the incremental height $\eta^h_{i,1}$ (plus one if there is another compatible proof $\omega^{h}_j$ with same history and $C(\omega^h_{i})=C(\omega^h_{j})$). We consider $b$-block confirmation rule where a ledger and the transactions inside are confirmed if the ledger is part of the heaviest chain and there are $b$ complete blocks mined on top of it.

Note that, as the fork choice rule is the chain that contains the most aggregate work at all times, this implies that the ledger offering the most fees is only a temporary fork choice rule. More specifically, when a block with $L$ proofs is created and the ledger offering the most fees is shared, honest miners try to mine an initiator proof for the next block height referring to this ledger. However, as soon as the next initiator proof is created, this proof contains the most aggregate work and is the fork choice rule no matter which ledger it refers to as long as that ledger is semantically valid. We consider such a situation in more detail in Appendix~\ref{app::fork_choice} and show that it does not create any splits in honest views.

\subsection{Reward Distribution}
The block at height $h$ receives total mining rewards proportional to the total proof height, $C(\omega^{h}_L)$, which is distributed in the ledger of block at height $h+1$. To mitigate the effect of the selfish mining attacks, if there are two proofs which have the same proof height, i.e., $C(\omega^{h}_i)=C(\omega^{h}_j)$ for some miners $i$ and $j$, then they split the reward of the proof height among themselves. In other words, a proof $i$ for block height $h$ receives $\frac{1}{C(\omega^{h}_L)}$ of the total block reward if it is the sole proof at that proof height, i.e., $C(\omega^{h}_i)\neq C(\omega^{h}_j)$ for any $j\neq i$, on the other hand if there is a parallel proof, each gets $\frac{1}{2C(\omega^{h}_L)}$ of the total block reward. 

Note that each proof in a cluster has the same difficulty contribution to the chain despite some might get half the reward for it (a punishment for parallelness). Thus, in total, the block at height $h$ receives mining rewards proportional to the total proof height, $C(\omega^{h}_L)$, which is distributed between the miners of the block at height $h$.

\subsubsection{Example Proof Model} The block at height $h$ in \figref{fig::our_parallel_protocol} receives $4$ units of rewards and each proof in this block gets $\frac{1}{2}$ units of rewards except the proof with incremental height $3$, which is the sole proof at proof height $2$ and gets $1$ unit of rewards. On the other hand, the block at height $h+2$ gets $7$ units of rewards and each proof gets $1$ unit of rewards.

\subsection{Adversarial Ledgers}\label{sec::adv_ledgers}
To have a valid ledger, we need  the ledger associated with $\omega^h_{i,1}$ of the initiator proof $w^h_i$ to be valid, i.e., the transaction ledger that an initiator proof accepts for the previous block height needs to be semantically correct with respect to the complete ledger history. Notice that, even if the other proposed ledgers of block height $h-1$ are invalid, as long as the associated PoW solutions are valid, they can be used by an initiator proof of block height $h$ as valid votes in the summary section, i.e., $\omega^h_{i,2}$. In other words, as long as $f_H(\omega^{h-1}_{j})<T_L$, then $\omega^{h-1}_{j}$ can be used in $\omega^{h}_{i,2}$ even if the associated proposed ledger is not valid or withheld. 
 
After the aggregate work at a cluster of block height $h-1$ reaches $L$ votes, the ledger of the proof that offers the most fees is expected to be shared with the network within a reasonable time (grace period) such as maximum network delays $\Delta+\delta$. Here, $\Delta$ is the maximum network delay when sending a transaction ledger of a block, whereas $\delta$ is the maximum network delay of sending a proof. If the ledger that offers the most fees is not shared within the grace period, the ledger offering second most fees (and the others if necessary) is shared. Moreover, in this case, the proof associated with the ledger that was not shared can still be included as a valid vote and its aggregate work is counted towards the heaviest chain.

To decrease the effect of adversarial ledger not being shared in a grace period which slightly delays the transition to the next height, we could allow a bit more than $L+1$ proofs, where the reward split for extra proofs could be omitted as well if the ledger offering most fees is not shared. We leave these minor adjustments to specific protocol implementations. Note  that, as it is common in the literature and the practice, we assume honest miners echo the messages they receive, hence all honest miners eventually receive the adversarial ledger if one of them receives it.

\section{Analysis of the Proposed Protocol}\label{sec::analysis_of_our_protocol}
In this section, we analyze the superior properties our protocol inherits from the existing PoW protocols \cite{btc-whitepaper,PHANTOM_GHOSTDAG,bitcoin_ng,decor+}. We start analyzing the resilience against double-spending, i.e., the consistency. Then, we consider the proof withholding attacks and show the resilience using an MDP analysis. We also consider the fairness of the transaction fee distribution rule of our protocol.

\subsection{Consistency Attacks}
For the purpose of analyzing the double-spending probability, here we make same assumptions about the network model and the miners. Our assumptions resemble the state-of-the-art analysis of double-spending in bounded delay models for Nakamoto consensus \cite{nakamoto-always-wins, guo-btc-sec-lat, our-sec-lat-extended, cao2023tradeoff, our-random-delay}. As analysis of double-spending probability in the bounded delay models requires a good understanding of the possible adversarial attacks within the delay intervals, we assume sufficient familiarity with the models/attacks/approaches in the cited blockchain analysis papers. Here, we try to explain the analysis in as simple and concise terms as possible. Note that, by showing that the upper bound on double-spending decreases exponentially, we essentially prove both the safety and the liveness properties under $\Delta$-synchronous settings due to the occurrence of Nakamoto blocks \cite{nakamoto-always-wins}.

\subsubsection{System Model} 
Assume that there are $n$ honest miners ($n$ is large) connected with an underlying network for communication, each holding an infinitesimal hashpower. Further, each proof generated by an honest miner experiences a network delay of $[0,\delta]$, whereas each block ledger experiences a network delay of $[0,\Delta]$. A single colluding adversarial miner is present in the system holding $\alpha$ fraction of the total hashpower. This adversarial entity experiences no communication delay and is aware of any proof as soon as it is generated. The adversary can delay the transmission of the honest proofs/ledgers by maximally allowed $\delta$/$\Delta$ for its own advantage and the ties are broken in the favor of the adversary.

\subsubsection{Confirmation Rule} 
Here, we only consider the double-spending probability for parameters $L\geq 10$, $b=1$ and give an upper bound for it. To recap, our $b$-block confirmation rule states the following: A ledger together with the transactions inside are confirmed if it is the part of the heaviest chain and there are $b$-complete blocks mined on top of it. Notice that, in Nakamoto consensus, a $b$-block confirmation rule also counts the block that contains the ledger, therefore, when we compare our results, we should take this difference into account. More formally, our $b=1$ rule corresponds to $1\leq b \leq2$ in Nakamoto consensus, since the ledger is proposed in one block and gets voted in the next (even though it remains hidden in the former block).

We refer the reader to Appendix~\ref{app::double_spend} for the description of the double-spending attack and the related analysis which relies on the state-of-the-art techniques for Nakamoto consensus; please see also \cite{nakamoto-always-wins, guo-btc-sec-lat, our-sec-lat-extended, cao2023tradeoff, our-random-delay}.

\begin{figure}[t]
    \centerline{\includegraphics[width=0.7\columnwidth]{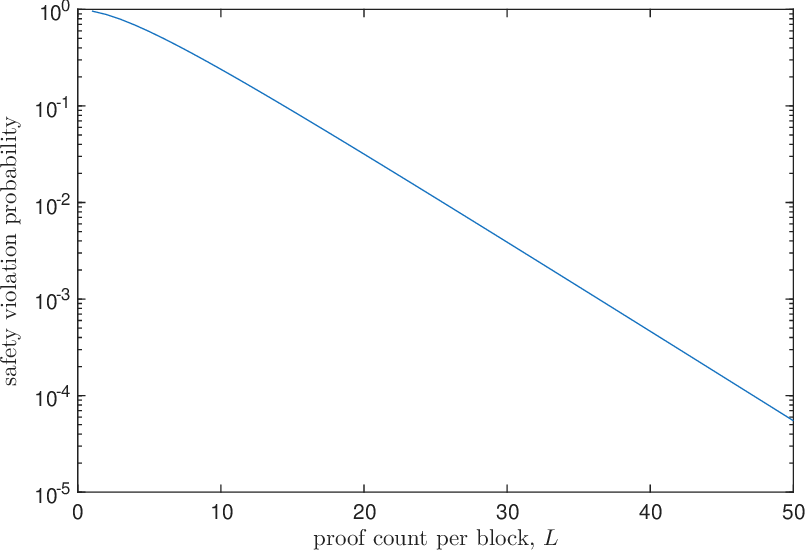}}
    \caption{$1$-block confirmation rule with $L$ proofs, $\alpha=0.25$.}
    \label{fig::consistency}
\end{figure}

\subsubsection{Numerical Evaluation}
To illustrate the strength of our protocol in the Bitcoin regime, we pick the following parameters for numerical evaluation of our theoretical results. We pick $\lambda_B=\frac{1}{600}$, $\Delta=10$ seconds \cite{DSN-Bitcoin-Monitoring} as usual for Bitcoin and $\alpha=0.25$ fraction of adversarial power. We further pick $\delta=1$ second. Notice that, $\delta=1$ second upper bound is on the delay of the transmission of a proof that has less than $1KB$ size is extremely pessimistic. Even though we assumed $L\geq 10$, to display the behavior of the curve fully we plot the double-spending curve for $L\geq 1$ in \figref{fig::consistency}. It is easy to verify that our protocol achieves an upper bound of double-spending probability around $10^{-4}$ before $L=50$. 

Note that the consistency guarantee of our model is similar to that of parallel PoW protocols \cite{parallel-pow-bounds}. This is an expected result, since by requiring multiple proofs for a block, whether parallel or sequential or a mix of them as ours, we decrease the chance of the adversarial ledger to be lucky with its attack and ahead of honest miners when the honest ledger is confirmed. This is simply a reduction of variance argument, which is in essence no different than increasing $b$-block confirmation rule in Bitcoin.

The results indicate a considerably superior security guarantees compared to Nakamoto consensus if we assume the same block sizes, generation times and network delays. More specifically, with $1$-block confirmation rule and $10$ minute block interarrival time, under $L=50$ and $\alpha=0.25$, an upper bound of double-spending probability around $10^{-4}$ is reached within less than $2$ blocks time in our protocol. In comparison, Bitcoin achieves a double-spending probability around $10^{-3}$ in $22$ blocks time\cite{our-sec-lat-extended}.

\subsection{Incentive Attacks}
Next, we consider the incentive attacks on our protocol. Mainly, there are two different incentive attacks, one where the adversary creates a block with all the proofs being adversarial and the other where the adversary hides its proofs for the current height and combines them with honest proofs of the current height to create a block. 

The analysis of the former attack is similar to our consistency analysis with $b=1$, which shows that the attack is less profitable than honest mining as $L$ increases. For example, even if we ignore the potential adversarial block losses in pre-mining and post-confirmation phases but take the probability of adversarial advantage of $M_{lead}$ and $M_{deficit}$ into account, the success probability of replacing $1$ honest block is with probability less than $10^{-4}$ for $L=50$ and $\alpha=0.25$. Notice that, if the attack fails, adversary has no proofs in the honest block, hence its relative rewards is upper bounded by the success probability of the attack, which is less than the relative rewards of honest strategy, i.e., $\alpha=0.25$, for $L\geq 10$.

We believe, it is trivial to do such an analysis, as it is done via Monte Carlo simulations in \cite[Section~V, Fig.~7]{bobtail}, which essentially shows that PoW protocols become more resistant to entire block withholding as the number of proofs required for a single block increases; hence, we omit such trivial confirmation of expected results here.

\subsubsection{Proof Withholding} 
The latter attack is what we call proof withholding attack in this paper and it is generally much harder to prevent in parallel PoW protocols. The analysis and discussion we provided in Section~\ref{sec::prev_parallel_pow} show that the existing parallel PoW protocols fail to prevent such attacks for even small values of $\alpha$ and they perform worse than Nakamoto consensus in general. Our protocol on the other hand, is a hybrid protocol where a block consists of multiple proofs like parallel PoW \cite{bobtail, tailstorm,dag_parallel_pow} 
and multistage PoW protocols \cite{multistage_pow, multistage_pow_props_vulnerabls, serial_vs_parallel_pow}, however the order of these proofs are semi-sequential, i.e., we allow at most two parallel proofs at a proof height which is inspired by the $k$-cluster rule in PHANTOM protocol \cite{PHANTOM_GHOSTDAG} with $k=1$. 

More specifically, since the proof propagation times are small, if we consider the DAG graph created by the proofs, for any proof within a block, at most one proof is allowed to be not an ancestor or descendant. Although this condition is stricter compared to DAG-style voting of \cite{dag_parallel_pow}, since the proofs we consider are small-sized and only refer to the ledger through the Merkle root, it is satisfied in most network conditions. Thus, the resilience against proof withholding attacks is increased compared to them while multiple proofs are still allowed for a block. Further, a punishment rule, i.e., reward splitting is introduced to further solidify the resilience inspired by the DECOR+ \cite{decor+} and RS protocols \cite{preneel_common_metrics}.

\subsubsection{Numerical Evaluation} 
\begin{figure}[t]
    \centerline{\includegraphics[width=0.7\columnwidth]{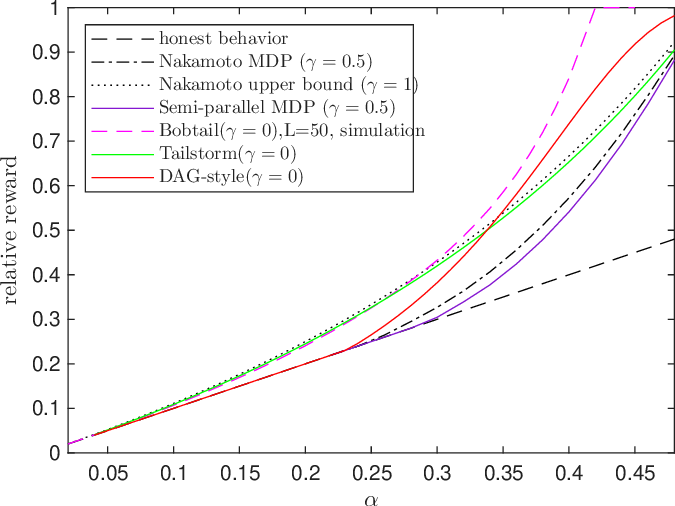}}
    \caption{Relative revenues of withholding attacks.}
    \label{fig::all_withhold_attacks}
\end{figure}

In Appendix~\ref{app::mdp}, we explain the MDP model with the appropriate state and action space as well as the objective function for our protocol. In \figref{fig::all_withhold_attacks}, we present the results of the MDP models together with the results of Section \ref{sec::prev_parallel_pow} where we analyzed the relative rewards of the existing parallel PoW protocols. We pick $\gamma=0.5$ for the MDP models and truncate the results of Section \ref{sec::prev_parallel_pow} before they rise above the honest behavior for the sake of neatness in the figures. For Nakamoto consensus, we run the MDP model of \cite{optimal-selfish} and the modified model described in in Appendix~\ref{app::mdp} is run for our protocol with $10^{-5}$ precision where $\alpha$ values range from $0.02$ to $0.48$ with a $0.02$ spacing between them. 

Note that the effect of $\gamma$ is ignored in the analysis of Section \ref{sec::prev_parallel_pow} for Tailstorm and DAG-style voting, which implies that the actual relative rewards for the adversary is underestimated and the best possible rewards are even higher than the values we show here. Nevertheless, it is clear that our protocol performs better even when the adversary is underestimated in other protocols. 

In our protocol, matching always gives a half reward when $(a,h,fork)=(1,1,relevant)$ (see Appendix~\ref{app::mdp}) whereas the adversarial reward in Nakamoto consensus increases with increasing $\gamma$ since matching succeeds more often. Thus, in comparison to the Nakamoto consensus, our protocol is more resilient against selfish mining attacks when $\gamma\geq0.3$ and resilience increases even more as $\gamma$ increases further, which we plot in Appendix~\ref{app::mdp}. On the other hand, when $\alpha>0.28$ as $\gamma$ drops below $0.3$, Nakamoto consensus becomes slightly more resilient. For the relative ratio in other values of $\gamma$, we refer the reader to Appendix~\ref{app::mdp}.

\subsubsection{Reward/Punish} 
We note that, according to \cite{Grunspan_witholding_resilience}, the orphan rewarding mechanisms make sure that the honest strategy is the best strategy in terms of the profitability, however, such an analysis ignores the dynamics of rational but honest miners who are free to leave when they are making more losses as explained in Appendix~\ref{app::relative_reward}. In our algorithm, we still resort to the orphan reward mechanism by rewarding parallel proofs. However, inherently, the goal in parallel PoW is to allow multiple proofs mined concurrently. Hence, it is not fair to reward one more than the other when two parallel proofs exist. Further, allowing orphans from past into the main chain makes the parallel PoW protocols vulnerable to proof withholding attacks. Thus, we pick reward splitting mechanism as a compromise. 

Notice that the authors of \cite{preneel_common_metrics} acknowledge the dilemma of rewarding the bad or punishing the good in a protocol design. They show that punishment protocols such as RS achieve better resilience against double-spending and selfish mining attacks when $\gamma=0.5$ but they suffer more against the feather-forking attacks \cite{feather-forking}. However, the analysis of resilience against a censorship attacker in our protocol is different than the protocols considered in \cite{preneel_common_metrics} since a feather-forking attacker would not be able to see the content of the transaction ledger associated with a proof until $L$-proofs are mined. Until then, the attacker will contribute to the same proof DAG as honest miners and it only mounts the attack after $L$ proofs are mined and if the ledger with maximum fee contains the transaction that attacker tries to censor. We do not analyze this type of attack here.

\subsection{Transaction Fee Split Rule} \label{sec::transaction_fee_split_rule}
The transaction fees associated with a ledger at block height $h-1$ is divided between the miner who proposed the ledger at block height $h-1$ and the miners of block height $h$ who voted for the ledger. The ledger creator (leader) receives $r$ fraction of the fees and the voters share $(1-r)$ fraction as a whole, which are divided between them in the same way coinbase rewards are divided, i.e., each proof height gets $(1-r)\frac{1}{C(\omega_L^h)}$ fraction of the total fees and is divided between two proofs if that proof height contains parallel proofs.

In our protocol, we choose $r=\epsilon$, i.e., the leader gets a tiny fraction of the transaction fee rewards. Notice that, when $L>1$, the transaction inclusion attack considered in \cite[Section~5.1]{bitcoin_ng} becomes different in our protocol. Specifically, when multiple miners have a proposed ledger at block height $h-1$, if the adversary offers the most fees and it does not share the ledger, another ledger will be shared after the grace period ends and the adversary loses the potential profits which implies as long as $r>0$, the adversary should publish its ledger if it offers the most fees. If an honest miner offers the most fee, the adversary still might try to mine on its own ledger.

In Appendix~\ref{app::fee_split_attack}, we consider such an attacker that acts against the rule of picking the ledger offering the most fees to increase its expected transaction fee revenue. We show that, as long as $r$ is small enough, such an attack is usually not profitable. Moreover, even if profitable, the effect on the transaction fee rewards received by honest miners is minimal. 

\section{Relation to Existing Works and Limitations} \label{sec::discussion}
Our protocol can be seen as a hybridization of sequential PoW mechanism with parallel PoW protocols. More specifically, in our protocol, each proof is both a vote for the previous block height and a proposal ledger for the current block height to be voted in the next block. However, the ledger is hidden under a hash summary which decreases the proof propagation times which in turn decreases the extent of parallelness needed in the protocol. Thus, we consider $k$-cluster rule of \cite{PHANTOM_GHOSTDAG} with $k=1$ and apply it to the parallel proofs of a block. In \cite{PHANTOM_GHOSTDAG}, a subset of blocks in a DAG is called $1$-cluster if for every block in the subset, all other blocks except one are either an ancestor or predecessor. Then, for any block, the block that is not an ancestor or predecessor has to be a parallel block in the sense that it should have the same ancestors and predecessors. 

The role of $k$ in \cite{PHANTOM_GHOSTDAG} essentially comes from the fact that multiple blocks can be generated during a block delay which are non-sequential blocks and would not be accepted in the original Nakamoto consensus. Thus, the authors of \cite{PHANTOM_GHOSTDAG} propose a modification to allow $k$ non-sequential DAG blocks. In our protocol, we pick $k=1$ since instead of blocks, the proofs are propagated through the network and the size of each proof is very small. 

On the other hand when the $L$th proof is generated, a block ledger is propagated through the network where the pick of $k=1$ might be too small for that specific proof height if the throughput is set high. This could be readjusted in the protocol by allowing more than $2$ parallel proofs in the last proof height. This in turn would require a different reward split rule for the last height. In this case, their ledgers could be ignored as well in order to prevent ledger-replications. Here, we assume that, such an event is negligible and leave it to the specific designers to consider these situations according to their throughput choice.

Note that, completely ignoring parallel proofs or accepting but not rewarding them would be against the original intention of creating a parallel PoW protocol that is fair and has less variance and allows proofs to be mined in parallel. On the other hand, allowing too many parallel proofs allows the adversary to mount withholding attacks as considered in Section~\ref{sec::prev_parallel_pow}. Moreover, it is also unnecessary to allow too many parallel proofs if proof sizes are very small that can be propagated through the network fairly quick as in our protocol. Thus, our pick of $k=1$ is a compromise between them. Future work could consider fair rewarding mechanisms while allowing higher $k$-cluster proof DAGs, where $k$ can be seen as a measure of parallelness. 

From the perspective of incentive mechanism, \cite{decor+, Publish_or_Perish, preneel_common_metrics} consider rules for Nakamoto consensus that increase the resilience against selfish mining attacks where the protocols considered in these works are very similar to $1$-cluster DAG of \cite{PHANTOM_GHOSTDAG}, hence we pick similar punishment mechanisms as considered in those papers. 

From the perspective of the throughput increase, our protocol is much less restricted compared to the existing work since the ledger is separated from the proofs. As a result, the ledger size can be increased arbitrarily with two things to bear in mind. First, as ledger size is increased, $\Delta$ increases which in turn leads to higher number of proofs that are mined during a grace period. These proofs should be treated differently than the parallel proofs on other proof heights and rewarded fairly, which we leave to specific implementations of designers. Second, although the ledger is created by a single miner for each block, which prevents transaction conflicts, when the ledger size is increased, a ledger creator gets more control of the transactions picked for a ledger. However, the effect of this is mitigated by the fact that the ledger offering the most fees is picked which prevents the ledger creators to act at their own pleasure when creating a ledger.

Although we analyzed incentive attacks on the coinbase rewards as well as the transaction fees individually, there is still a need to develop an analysis considering both attacks at the same time to give stronger guarantees on the resilience against incentive attacks. However, we note that, even for Nakamoto consensus, which has been thoroughly studied for over fifteen years, such a rigorous analysis is still missing \cite{selfless_sok}.  
 
\appendices

\section{Bobtail Smallest Hash Attacks}\label{app::bobtail}
 
\subsection{DoS Attack in Bobtail} 
If an attacker has the lowest hash valued proof which is published together with $L-1$ other valid proofs mined by everyone, it can simply refuse to publish the block, i.e., Denial-of-Service (DoS) attack. This attack, acknowledged in \cite[Section~VIII.C]{bobtail}, requires honest miners to further mine until finding a lower hash value to create a block.

\subsection{Proof Withholding in Bobtail}
The authors of Bobtail also consider Monte Carlo simulations of proof withholding attacks where a miner simply withholds all its proofs for the current block height until it finds the lowest hash valued proof or gives up the attack and releases all its proofs if it seems unlikely to do so before honest miners creating a block with all proofs honest \cite[Section~VIII.B]{bobtail}. However, the authors do not consider that if the adversary manages to mine the proof with the lowest hash value for the current block height, it can release this lowest hash proof while withholding the rest of the proofs, to build a lead for the proofs of the next blocks. This in turn becomes similar to the DoS attack described earlier, since honest miners will have to find a proof with lower hash value. We describe this hard-coded attack next and provide simulations to display the effectiveness of the attack.

\subsection{DoS Proof Withholding Attack}
Let us assume that everyone works on the same block and $L$ proofs are needed. When honest miners create a proof, they release immediately, whereas the adversary keeps its proofs hidden, however, while picking a support value, it acts honestly, i.e., whenever it mines a proof, it simply picks the lowest hash value among all proofs mined so far as the support value, no matter if that proof is mined by honest miners or the adversary itself. When the system has $L$ proofs in total, the adversary (which produced $L\alpha$ proofs on average) checks if it has the lowest hash proof among all $L$ proofs:
\begin{enumerate}
    \item If the adversary does not have the lowest hash value, it simply releases all its proofs and the honest miner who has produced the lowest hash value creates the block including all the $L$ proofs. The adversary does not lose any primary or bonus reward described in \cite[Section~VIII]{bobtail}.
    \item \label{enum::bob_attack_2nd}If the lowest hash valued proof belongs to the adversary, it releases this proof, i.e., makes it public, while keeping the remaining proofs ($L\alpha-1$ proofs on average) hidden. At that point, honest miners see $L(1-\alpha)$ honest proofs on average plus $1$ adversarial proof which has the lowest hash value. They keep mining to complete $L$ public proofs since they are not aware of the hidden adversarial proofs, where, they use the adversarial proof as support until they mine a proof with lower hash value. The adversary, on the other hand, moves on to mine the next block to build a lead. 
    
    When $L$ public proofs are completed ($L-1$ honest and $1$ adversarial):
    \begin{itemize}
        \item If the honest miners already produced a proof with lower than the public adversarial proof, they can create a block. To prevent that, the adversary releases all its work for that block plus some additional work from the lead that it has created for the next block to undo all honest work. It simply loses no reward and undoes some honest work.
        \item If the public adversarial proof is still the lowest hash value, the honest miners have to waste their time on the current block until they find a lower hash, while the adversary increases its lead further on the next block. It will simply undo the honest work when honest miners find a lower hash value. It simply loses no reward and undoes some honest work. 
    \end{itemize}
\end{enumerate}

\begin{algorithm}[h]
    \caption{Bobtail DoS-Proof Withholding Attack}\label{alg::bobtail_attack}
    \begin{algorithmic}[1]
    \WHILE{$attack$}
    \STATE $cur\_blck\_adv\_cnt$\\$\coloneqq  initial\_released\_lead+still\_hidden\_lead()$
    \STATE $cur\_blck\_tot\_prf\_cnt$\\$\coloneqq cur\_blck\_adv\_prf\_cnt$+$cur\_blck\_hon\_prf\_cnt$
    \STATE $mine\_hidden\_until\_L\_total\_proofs()$
    \IF{$is\_smallest\_hash\_adversarial()$}
        \STATE $release\_smallest\_hash\_prf()$
        \WHILE{$is\_smallest\_hash\_still\_adversarial()$}
        \STATE $mine\_lead\_for\_next\_block()$
        \ENDWHILE
        \IF{$is\_lead\_enough\_to\_undo\_honest\_work()$}
        \STATE$release\_cur\_adv\_block()$
        \STATE$release\_initial\_lead\_for\_next\_block()$
        \ELSE
        \STATE $adv\_loses\_rewards\_for\_cur\_block\_and\_lead()$
        \ENDIF
    \ELSE
    \STATE $release\_all\_proofs()$
    \ENDIF
    \STATE $continue\_next\_block\_loop()$
    \ENDWHILE
    \RETURN $relative\_reward\_ratio$
    \end{algorithmic}
\end{algorithm}

We also provide a pseudocode  for the attack in Algorithm~\ref{alg::bobtail_attack}. With this attack, none of the attack prevention mechanisms of Bobtail \cite[Rule I, II, Section~VIII.B]{bobtail} are valid since the system follows most aggregate work, which is the adversarial chain and it will undo the honest work, increasing the relative reward of the adversary. In fact, the bonus reward system introduced in \cite[Section~VIII]{bobtail} will even worsen the situation for honest miners. If honest miners mine the lowest hash proof, the adversary gets the bonus reward as it acts honestly when referring to the lowest hash proof. If the adversary mines the lowest hash proof, honest miners become aware of the lowest hash valued proof after there are already $L$ proofs in the system as the adversary keeps private before that point, hence honest miners lose the bonus rewards.

Note that, the adversary has access to all public information, can track the public work and thus avoid the risk of losing its hidden work. This attack results in higher profits for the attacker even for small $\alpha$, $\gamma$ values and the profitability threshold is $\alpha_T=0$. The reasoning is no different than the selfish mining attack on smallest-hash tie-breaking (SHTB) rule which has profitability threshold of $\alpha_T=0$. The authors of \cite{preneel_common_metrics} show that the informational advantage an attacker has in SHTB protocols allows it to gain more rewards than honest mining by withholding attacks in smallest hash regions. Same arguments hold here which shows that the profitability threshold is $\alpha_T=0$, which is worse than Nakamoto consensus. 
 
Notice that, to succeed further in this attack, the adversary can further modify the attack to keep its proofs hidden (Step~\ref{enum::bob_attack_2nd}) only if the smallest hash it has produced is below a certain threshold. It could simply minimize its risk by picking a low enough smallest hash region mentioned in \cite{preneel_common_metrics}; see \cite{preneel_common_metrics} to see why SHTB rule is a weak choice in consensus models against incentive attacks. 

For example, in \figref{fig::bobtail}, if $p_5$ is mined by the adversary and the adversary is confident that the hash value of $p_5$ is low enough (in smallest hash region), it can refuse to share the block it creates and it can safely mine for the next block before honest miners can do so for a while, gaining an advantage. Although the authors state that DoS attacks can be thwarted by banning the attackers, such a defense mechanism is not straightforward in permissionless blockchain settings with easy-to-obtain sybil identities.

We first provide some theoretical intuition for the attack and then explain the simulation environment. Notice that, the first time there are $L$ proofs in total, i.e., counting both the adversarial and honest proofs, with probability $\alpha$, the lowest hash proof belongs to the adversary. In other words, initially, the attack in Step~\ref{enum::bob_attack_2nd} happens on average in $\frac{1}{\alpha}$ blocks. Once this happens, the fraction of adversarial proofs for the next block will increase since the adversary mines a lead for the next block before honest miners can find a smaller hash for the current block. Considering the value of the hashes are essentially numbers in bits, which can be represented as uniform distribution on $[0,1]$, this attack actually enters in vicious cycle where the adversary has practically infinite lead. 

To understand why, consider the distribution of the minimum value of $L$ i.i.d. uniform distributions., which has the following pdf and cdf
\begin{align}
    f(x)&=L(1-x)^{L-1},\\
    F(x)&=1-(1-x)^L.
\end{align}
Let $H_{waste}$ denote the number of honest proofs that has to be created until the honest miners hit a lower hash value. Now, assuming the adversary has created the lowest hash value $x$, 
\begin{align}
    \mathbb{E}[H_{waste}|x]&=\frac{1}{x}.
\end{align}
Now, considering the distribution of $x$, we have
\begin{align}
    \mathbb{E}[H_{waste}]&=\int_{0}^{1} L\frac{(1-x)^{L-1}}{x} \,dx\\
    &=L\left[\ln x+\sum_{n=1}^{L-1}\binom{L-1}{n}(-1)^n\frac{x^n}{n}\right]^1_0,
\end{align} 
which diverges. This essentially means that the adversary is able to hit so low hash values that the honest miners have to waste virtually infinite amount of energy to replace it. From a practical perspective, if the block is not created after the first $L$ public proofs, honest miners could start over with ignoring the adversarial lowest hash proof. In this case, they will start to mine until they create $L$ new proofs or a proof with lower hash than the adversarial, whichever comes first. Thus, $H_{waste}$ becomes essentially the expectation of a truncated geometric distribution capped at $(L+N)$, where $N$ denotes the number of hidden adversarial proofs that were created for the current block before the adversary started to build a lead for the next block. For the first block the adversary gets the lowest hash proof, we have
\begin{align}
    \mathbb{E}[H_{waste}]&=\int_{0}^{1} L(1-x)^{L-1}\times\left(\sum_{k=1}^{L+N-1}k(1-x)^{k-1}x+(L+N)(1-x)^{L+N-1}\right) \,dx\\
    &=L\sum_{k=1}^{L+N-1}\frac{k}{(L+k)(L+k-1)}+\frac{L(L+N)}{2L+N-1}.
\end{align} 
If we assume $N\approx L\alpha-1$, with $L\geq10$, this in turn results in more than $\frac{2L}{3}$ honest proof waste for $\alpha>0$. Meanwhile, the adversary creates the proofs for the next blocks, increasing the fraction of the adversarial proofs for the next blocks, which in turn can increase average honest waste further for the following blocks if the adversary gets the lowest hash again. 

We leave a more detailed theoretical analysis that accounts for the lead in the next blocks to the reader if interested, and move on to simulate the attack with the rule where the honest miners start over with ignoring the adversarial lowest hash proof if the adversary has it and refuses to publish a block.

\begin{figure}[t]
    \centerline{\includegraphics[width=0.7\columnwidth]{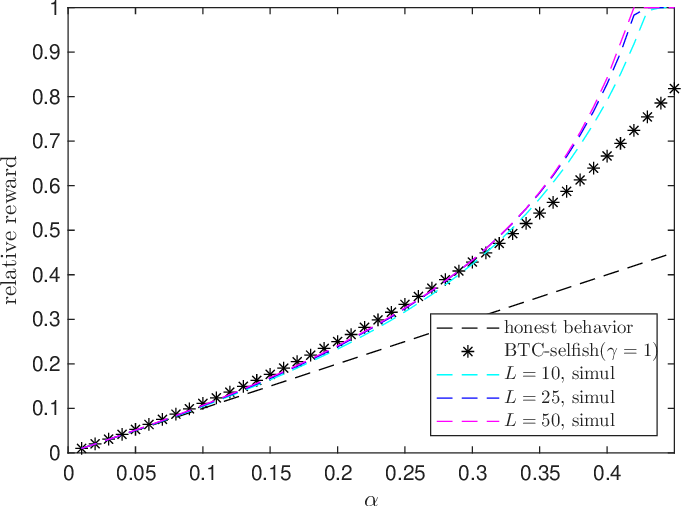}}
    \caption{Bobtail relative reward of DoS proof withholding attack.}
    \label{fig::bobtail_simu}
\end{figure}

Results with the practical rule is presented in \figref{fig::bobtail_simu}, where we simulate the attack for $10^6$ blocks for $L=10,25,50$. Even with the extra protection that we add with this rule, it is clear that the profitability threshold is zero and the adversarial relative reward is at par with the upper bound of selfish mining attack in Nakamoto consensus, i.e., when $\gamma=1$. As $\alpha$ rises above $\frac{1}{3}$, the attack becomes even more powerful than the upper bound of selfish mining attack in Nakamoto consensus. Moreover, without the practical rule we introduced, our simulations do verify our theory that the explosion happens when we simulate a sufficiently large number of blocks.

\section{Why Relative Reward Metric?}\label{app::relative_reward} 
To understand why relative reward metric is the correct choice and why the normalized reward metric of \cite{tailstorm,dag_parallel_pow} is an illusion, consider the adversarial (honest resp.) miners who hold $\alpha$ ($1-\alpha$ resp.) fraction of the hashpower in the system. Let the total adversarial (honest resp.) reward per chain progress be denoted as $\mu$ ($\nu$ resp.). Clearly, an incentive compatible system that is not under an attack, should satisfy
\begin{align}
    \frac{\nu}{1-\alpha}=\frac{\mu}{\alpha},
\end{align}
which is an equality between the rewards per chain progress per computational power of the adversarial and honest miners.

When the adversary is employing an incentive attack, let $\mu_{attack}$ and $\nu_{attack}$ denote the new rewards per chain progress and let $\rho$ denote the fraction of the total rewards the adversary gets. Assume that the adversary gets more relative rewards than its fair share, i.e., $\rho>\alpha$, but the total adversarial reward per chain progress decreases compared to not employing the attack i.e., $\mu>\mu_{attack}$. 

We will analyze this system in more detail. However, to give the gist of our argument in a few words, if such a system that is under attack is safe as claimed in \cite{tailstorm,dag_parallel_pow}, then honest mining should be sustainable even if the adversary launches the attack. However, if honest mining is sustainable under the attack given the mining rewards and costs, then the adversarial mining is not only sustainable but also more profitable since adversary gets more relative rewards compared to its costs.

Specifically, an incentive compatible system that is safe against the attack should have
\begin{align}
    \frac{\nu_{attack}}{1-\alpha}\geq \frac{\mu_{attack}}{\alpha},\label{eq::progress_relative}
\end{align}
which compares the rewards per chain progress per computational power which is not satisfied since
\begin{align}
    \frac{\nu_{attack}}{1-\alpha}<\frac{\nu_{attack}}{1-\rho}= \frac{\mu_{attack}}{\rho}<\frac{\mu_{attack}}{\alpha}.\label{eq::progress_relative_2}
\end{align}
Notice that, in this situation, if the adversary is getting less reward per chain progress in an incentive attack, \eqref{eq::progress_relative_2} implies that the honest miners are affected even worse since relative reward $\rho$ is in the favor of the adversary. In other words, under the attack, the normalized honest loss is larger, i.e., 
\begin{align}
    \frac{\nu-\nu_{attack}}{1-\alpha}> \frac{\mu-\mu_{attack}}{\alpha}.
\end{align}
As a result, if attacking is not profitable for the adversary despite getting higher relative reward than its fair share, i.e., despite $\rho>\alpha$, then, for honest miners the situation is worse. In an open economy model, if the honest miners are rational and they are making losses (greater than the adversary per computational power) and their mining costs are not compensated by the rewards, they will leave the system before the adversary, which breaks down the whole blockchain system. This result is even more severe than having a system that rewards miners unfairly. 

On the other hand, if the system is secure under the attacks (as \cite{tailstorm,dag_parallel_pow} claim) and honest miners are not leaving the system in an open economy model, it implies that the coin exchange value compared to their mining costs dynamically changed and honest mining became sustainable, which in turn, implies that the adversary is making unfair profits as it gets more than its fair share. In other words, under the attack, if the honest reward per chain progress per computational power, $\frac{\nu_{attack}}{1-\alpha}$ is sustainable when taking the mining costs into account, then adversarial reward per chain progress per computational power is sustainable if $\frac{\mu_{attack}}{\alpha}= \frac{\nu_{attack}}{1-\alpha}$. Since $\rho>\alpha$, the adversarial reward per chain progress per computational power is not only sustainable but also more profitable, i.e., 
\begin{align}
    \frac{\mu_{attack}}{\alpha}> \frac{\mu_{attack}}{\rho}>\frac{\nu_{attack}}{1-\alpha},
\end{align} where $\rho$, i.e., the relative reward metric, can be compared to $\alpha$ to measure this extra profit.

In short, if the total reward issued per chain progress is $\mu+\nu$, then, an incentive compatible system should satisfy $\mu=(\mu+\nu)\alpha$. If the total rewards is $c(\mu+\nu)$, then, an incentive compatible system should satisfy $c\mu=c(\mu+\nu)\alpha$. If the adversarial fraction of the rewards satisfies $\rho>\alpha$, the system is not incentive compatible. The specific value of $c$ does not matter, as the number of coins issued will not change the market cap of the system and only changes the value of a single coin. Further, from an economic perspective, minting more or less coins in a cryptocurrency environment is not relevant either, as eventually the share of the coins held by an agent determines its purchasing power, not the amount of coins minted \cite{mankiw_macroecon}. 

What \cite{tailstorm,dag_parallel_pow} do, is decreasing the value of $c$ when the system is under the attack and focusing on the loss of the adversary $\mu_{attack}$, ignoring a potentially larger and disastrous loss on the honest side $\nu_{attack}$ and claiming their system is secure. Since \cite{tailstorm,dag_parallel_pow} assume that the system is secure under the attacks, which implies that honest miners are rewarded enough to continue on mining, and then they should focus on the adversarial rewards to see if they get more than their fair share, i.e., check if \eqref{eq::progress_relative} is satisfied or not, which is equivalent to checking if $\rho\leq\alpha$ or not. 

That is what we do in this paper, i.e., we focus on the relative reward metric rather than focusing on adversarial reward per chain progress $\mu_{attack}$ that completely ignores the side of the honest rewards and honest mining costs which creates an illusion. 

\section{Proof Withholding Attack for Tailstorm and DAG-Style Voting}\label{app::witholding_attack}
We start by explaining the attack structure. Assume that the adversary withholds its proofs at the current block height, call it $h$, and combines them with the honest proofs of the current block height to create the block associated with these proofs but withholds it. In this situation, this adversarial block contains $\alpha$ fraction of the adversarial proofs on average. Next, it starts creating proofs for the next height, $h+1$, while honest miners keep mining on the previous height $h$ since the adversary did not share its own proofs for the block height $h$. When honest miners find the final ($L$th) proof to create a block at block height $h$, the adversary simply releases its block for height $h$, plus one additional proof for height $h+1$ to make sure the public part of its chain has the most aggregate work and some honest work are undone. Hence, everyone starts to create proofs for the block height $h+1$ on top of the public block at height $h$ released by the adversary recently.  Notice at this point, there is some initial advantage that the adversary has for proofs at height $h+1$ since it started mining on them as soon as it created the block at height $h$. The adversary keeps these proofs and the others it creates later private again until honest miners reach $L$ proofs for the block height $h+1$, at which point the adversary releases its block at height $h+1$ and one additional proof on height $h+2$ and keeps the rest of the proofs private again. The attack continues in a similar manner for the next blocks.

\begin{figure*}[t!]
\centerline{\includegraphics[width=1\columnwidth]{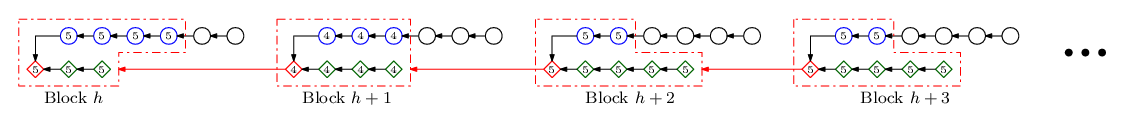}}
	\caption{Tailstorm withholding attack.}
	\label{fig::tailstorm}
\end{figure*}

In \figref{fig::tailstorm}, we visualize the proof withholding attack in Tailstorm protocol with $L=7$. The circles and rhombus are proofs created by the honest and adversarial miners, respectively, and the numbers inside represent the rewards they get. Black circles represent the honest work that do not end up in the longest chain due to the attack. At block $h$, initially only the red proof is created and shared by the adversary and honest miners create $4$ proofs (blue) which they share while the adversary creates $2$ more proofs (green) that it hides. Since honest miners are not aware of the $2$ adversarial proofs, they create $2$ more proofs (black). At this point, the adversary shares the green proofs at block $h$ together with the red proof at block $h+1$, which makes sure the black circled honest proofs are replaced in the longest chain. The attack continues in a similar manner for next heights. 

\begin{figure*}[t!]
\centerline{\includegraphics[width=1\columnwidth]{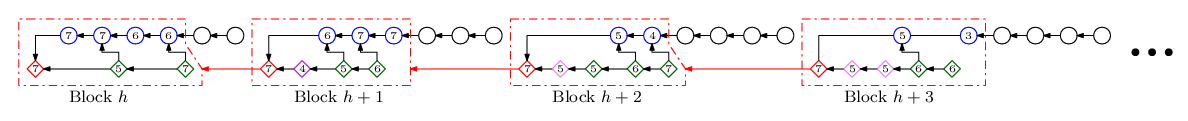}}
	\caption{DAG version of the withholding attack.}
	\label{fig::dag_2_attack}
\end{figure*}

We first prove the fraction of the adversarial proofs in the public chain in the long run under the attack.
\begin{lemma}\label{lemma::witholding_attack_limit}
    For both tree and DAG-style voting, in the long run, under the proof withholding attack, the fraction of the adversarial proofs in the public chain reaches $\frac{\alpha}{1-\alpha}\frac{L-1}{L}$  even when $\gamma=0$.
\end{lemma}
\begin{Proof}
    Let $a_n$ be the fraction of adversarial proofs in the public block at height $n$. Let $b_{n+1}$ be the fractional advantage of the adversary for block height $n+1$, i.e., the fraction of the proofs it already had for block at height $n+1$ when it released the complete block at height $n$. Then, according to the proof withholding strategy described above, the following holds,
\begin{align}
    \mathbb{E}[b_{n}|a_{n-1}]&=\frac{\alpha}{1-\alpha}\left(a_{n-1}-\frac{1}{L}\right), \label{eq::iterative_advantage_bn}\\
    \mathbb{E}[a_{n}|b_n,a_{n-1}]&=b_n+(1-b_n)\alpha=\alpha(1+a_{n-1})-\frac{\alpha}{L}. \label{eq::iterative_tot_an}
\end{align}
The reasoning is as follows: Given $a_{n-1}$, honest miners were not aware $a_{n-1}-\frac{1}{L}$ fraction of the proofs created by the adversary at height $n-1$. Note that, $1$ adversarial proof of the height $n-1$ was released initially to make sure the switch to adversarial chain happens, which explains $\frac{1}{L}$. Then, while honest miners try to mine $a_{n-1}-\frac{1}{L}$ proofs to create the block at height $n-1$, the adversary mines on height $n$. For every honest proof at height $n-1$, the adversary creates $\frac{\alpha}{1-\alpha}$ proofs at height $n$ on average, which explains \eqref{eq::iterative_advantage_bn}. Given $b_n$, $\alpha$ fraction of the rest of the proofs at height $n$ are mined by the adversary, which explains \eqref{eq::iterative_tot_an}. Clearly $\mathbb{E}[a_1]=\alpha$ which is for the first block where the attack started, which implies
\begin{align}
    \lim_{n\to\infty}\mathbb{E}[a_n]=\frac{\alpha}{1-\alpha}\frac{L-1}{L}.
\end{align}
Hence, in tree- and DAG-style voting protocols, the fraction of the proofs created by the adversary reaches $\frac{\alpha}{1-\alpha}$ in expectation as $L\to\infty$, which is the worst case scenario of selfish mining in Nakamoto consensus, i.e., when $\gamma=1$ \cite{selfish-mining,optimal-selfish}. Note that, there can be better attack strategies. For example, the attack did not take advantage of $\gamma$, hence the adversary always released $1$ extra proof which it does not have to do for higher $\gamma$ values.  Thus, $\frac{\alpha}{1-\alpha}\frac{L-1}{L}$ is essentially a lower bound.
\end{Proof}

We note that the analysis above is valid for $\alpha\geq\frac{1}{L}$ with $L$ large enough. A more rigorous analysis is needed that has to take into account the cases where the adversary becomes extremely lucky and has $a_n>L$ for some block heights $n$ as well as extremely unlucky cases and has $a_n-1<\frac{1}{L}$ for some block heights $n$. Although these cases can slightly change the final result of the analysis, for the purposes and claims of our results, the above analysis that considers the expected values is sufficient.

With the hard-coded proof withholding attack, the fraction of the adversarial proofs in the public chain reaches $\frac{\alpha}{1-\alpha}\frac{L-1}{L}$ even when $\gamma=0$, for both tree- and DAG-style voting. Note that, in Tailstorm \cite{tailstorm}, although the reward a block gets is scaled by the depth of the proof tree, everyone is punished and rewarded equally. Hence, the fraction of adversarial proofs we derive, is the fraction of the rewards the adversary gets. We note that the authors of \cite{tailstorm} consider a protocol variant, TS/const, where the relative reward of the adversary equals reward per chain progress. Their simulations include an incentive attack called minor delay against TS/const, which seems similar to the attack we describe here. However, the claims of the authors in \cite[Figure~4]{tailstorm} regarding the attack are not verifiable given our theoretical analysis. Hence, in Appendix~\ref{app:withold_sim_tree_dag}, we provide simulations of a practical implementation of the attack for $10^5$ blocks and verify our theoretical results.

\subsection{DAG Version of the Attack}\label{app::dag_withholding}
The fraction of the adversarial proofs under the attack does not equal the relative rewards of the adversary in DAG-style voting since we need to count the number of ancestors and descendants for each proof. To make the explanation easier to follow, we refer to \figref{fig::dag_2_attack} whenever suitable. The colors and shapes have the same meaning as in \figref{fig::tailstorm}, but we additionally have violet colored adversarial proofs, that indicate that those proofs were mined before the red proof was shared. The number inside the proofs represent the reward each proof gets. We note that, in the DAG version of the attack, the adversary points to all other adversarial proofs and all public honest proofs even though it does not release its own proofs immediately.

In short, for each block, the red proof is the advantage proof shared initially by the adversary and seen by everyone and hence gets full reward. All adversarial proofs get rewards proportional to the total number of adversarial proofs in the block. The greens (adversarial) additionally get rewards for the honest proofs they can see when they were mined. Blue proofs (honest) also get rewards for each green proof that refers to them in addition to the total number of blue and red proofs.

\begin{algorithm}[h]
    \caption{Practical Withholding Attack for \cite{dag_parallel_pow,tailstorm}}\label{alg::withhold_attack}
    \begin{algorithmic}[1]
    \WHILE{$attack$}
    \STATE $cur\_blck\_adv\_cnt$\\$\coloneqq  initial\_released\_lead+still\_hidden\_lead()$
    \STATE $cur\_blck\_tot\_prf\_cnt$\\$\coloneqq cur\_blck\_adv\_prf\_cnt$+$cur\_blck\_hon\_prf\_cnt$
    \STATE $mine\_hidden\_until\_L\_total\_proofs()$
    \STATE $cur\_blck\_public\_prf\_cnt$\\$\coloneqq initial\_released\_lead+cur\_blck\_hon\_prf\_cnt$
    \IF{$cur\_blck\_hidden\_adv\_prf\_cnt>\frac{1}{\alpha}$} 
        \WHILE{$cur\_blck\_public\_prf\_cnt<L$}
            \STATE $mine\_lead\_for\_next\_block()$
            \ENDWHILE
            \IF{$is\_lead\_enough\_to\_undo\_honest\_work()$}
            \STATE$release\_cur\_adv\_block()$
            \STATE$release\_initial\_lead\_for\_next\_block()$
            \ELSE
            \STATE $adv\_loses\_rewards\_for\_cur\_block\_and\_lead()$
            \ENDIF
    \ELSE
    \STATE $release\_all\_proofs()$
    \ENDIF
    \STATE $continue\_next\_block\_loop()$
    \ENDWHILE
    \RETURN $relative\_reward\_ratio$
    \end{algorithmic}
\end{algorithm}

\begin{lemma}
    In DAG version of the proof withholding attack, the fraction of the adversarial rewards reach 
    \begin{align}
        \rho=\frac{\mathbb{E}[A_R]}{\mathbb{E}[A_R]+\mathbb{E}[H_R]},
    \end{align}
    where
    \begin{align}
        \mathbb{E}[A_R]=&1\cdot L+(L\mathbb{E}[b_n]-1)\cdot L\mathbb{E}[a_n]+L(\mathbb{E}[a_n]-\mathbb{E}[b_n])\cdot L\frac{1+\mathbb{E}[a_n]}{2},\label{eq::dag_attack_adv}\\
         \mathbb{E}[H_R]= &L(1-\mathbb{E}[a_n])\cdot(1+L(1-\frac{\mathbb{E}[a_n]+\mathbb{E}[b_n]}{2})).\label{eq::dag_attack_hon}
     \end{align}
\end{lemma}  

\begin{Proof}
The adversarial proof released initially to make sure the switch to adversarial chain happens, is seen by everyone and simply gets $L$ rewards. This is represented as the red rhombuses in \figref{fig::dag_2_attack}. The number of advantage proofs, i.e., the proofs that the adversary mined before honest miners finished the previous block height, equals $Lb_n-1$ according to Lemma~\ref{lemma::witholding_attack_limit}. These proofs are represented as the violet rhombuses in \figref{fig::dag_2_attack} and are seen by only the other adversarial proofs and do not refer to any honest proof. This is because, while they were being mined, honest miners were still working on the previous height. Hence, there were no honest proof at that block height to refer to when the violet proofs were mined. As a result, these proofs get $La_n$ rewards, i.e., the number of adversarial proofs at that height. 

It remains to deal with the rest of the adversarial proofs, colored green in \figref{fig::dag_2_attack}. It is easy to see that there are $L(a_n-b_n)$ of them. 
However, the reward they get is slightly more complicated than the other ones because we have to count how many honest proofs they can refer to, i.e., the number of honest proofs that were already mined at their creation time.  Notice that, the first green adversarial proof is likely to see almost no honest proofs whereas the last adversarial proof sees almost all honest proofs. As the mining time of the proofs are uniformly distributed, on average, each of these green adversarial proofs sees $L\frac{1-a_n}{2}$ honest proofs at their block height, i.e., on average half of the honest proofs at their height. Of course, they also see all the adversarial blocks, i.e., $La_n$ of them.

Combining the counts and rewards of each color mentioned above, where we multiply the expectation of the count of each proof color and their expected rewards, the reward of the adversary equals \eqref{eq::dag_attack_adv} on average from each block.

For the honest proofs, colored blue and their total count is simply $L(1-a_n)$, similar reasoning applies. Honest proofs, in addition to seeing each other, also see the red adversarial proof but no other adversarial proof. Thus, their reward coming from blue and red proofs is $1+L(1-a_n)$ in total. However, they are also referred to by green proofs hence also get rewarded from them. Note that, the first blue proof is likely to be seen by almost all green proofs, whereas the last one is likely to be seen by almost none. We can make the argument about uniform distribution of mining times again, hence, each honest proof gets average reward $\frac{L(a_n-b_n)}{2}$, coming from the green proofs that refers to it.

Combining the counts and rewards of each color mentioned above, the reward of the honest miners is equal to \eqref{eq::dag_attack_hon} on average from each block. 
\end{Proof}

Note that both $\mathbb{E}[a_n]$ and $\mathbb{E}[b_n]$ are derived in Lemma~\ref{lemma::witholding_attack_limit}.
  
\subsection{Simulations}\label{app:withold_sim_tree_dag}

\begin{figure}[t]
     \centering
     \begin{subfigure}[b]{0.48\textwidth}
         \centering    \includegraphics[width=\textwidth]{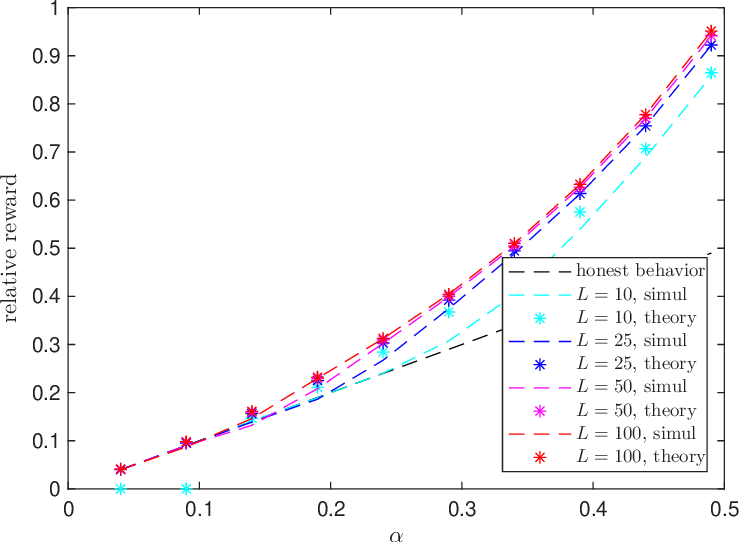}
         \caption{Tailstom version}
         \label{fig::tail_simu}
     \end{subfigure}
     \hfill
     \begin{subfigure}[b]{0.48\textwidth}
         \centering
    \includegraphics[width=\textwidth]{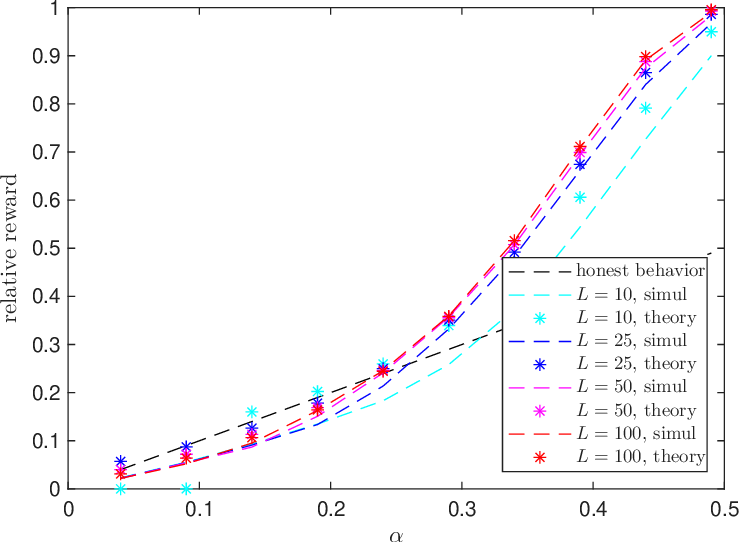}
         \caption{DAG version}
         \label{fig::dag_simu}
     \end{subfigure}
     \caption{Withholding attack simulation, $\gamma=0$.}
	 \label{fig::simulations}
\end{figure}

As the results claimed by \cite{tailstorm} are not compatible with our theoretical analysis, we simulate the attacks in both tree- and DAG-style voting where we apply some practical modifications that arise when $L$ is finite. The modification is as follows: The first time when there are $L$ proofs including the hidden adversarial ones for block $n$, the adversary releases all its proofs if the number of hidden proofs it has is not more than $\frac{1}{\alpha}$. This is because, in expectation, it takes $\frac{1}{\alpha}$ proofs to encounter an adversarial proof, which will be the initial lead for the next block that undoes the honest work. If the adversary has more than this, it is reasonable to take the risk, else, if it does not release the hidden proofs, it will simply lose them since there will be more newly mined honest proofs (black proofs in \figref{fig::tailstorm} and \figref{fig::dag_2_attack}). This situation happens because $L$ is finite and we have to deal with integers. 
The pseudo-code of the practical version of the attack is provided in Algorithm~\ref{alg::withhold_attack}.

The results are provided in \figref{fig::simulations}, where we run the simulation for $10^5$ blocks for each parameter. As we mentioned in our theoretical analysis, the accuracy of our analysis depends on $L$ as we use expected values. Hence, as $L$ increases, the theoretical and simulation values converge. Further, the attack does not need any network advantage $\gamma$ and both protocols become less resistant to incentive attacks than the original Nakamoto consensus. We note that, we did not dig through all possible ways to find better practical implementations for which the simulation results could converge faster even for smaller $L$. We leave this to interested readers, since our purpose is to show the vulnerability of the existing parallel PoW protocols.

\section{Fork Choice Rule} \label{app::fork_choice}
The fork choice rule is to pick the chain that contains the most aggregate work at all times. Thus, if $L$ proofs are mined for a block and the adversary owns the valid ledger offering the most fee but shares the ledger with only some of the honest miners in the grace period, these honest miners try to create an initiator proof of the next block height referring to the adversarial ledger while other honest miners try to create an initiator proof referring to another ledger. However, this is not a threat for consensus mechanism since as soon as someone creates and shares the next initiator proof, every honest miner will pick that newly created proof and the ledger it refers to as the fork choice since it contains the most aggregate work. In such a situation, by not sharing the ledger with all honest miners in time, the adversary risks any potential extra income due to owning the ledger offering the most fee since the first  initiator proof created next might refer to another ledger as the winning ledger. 

On the other hand, if an honest miner owns the ledger offering the most fee and the adversary creates the first initiator proof for the next block height, honest miners will accept the adversarial initiator proof since it is the chain with most aggregate work. Notice, the adversarial initiator proof might refer to another valid ledger instead of the ledger offering the most fee. Such a situation does not create any issue from the perspective of the consensus since the fork choice rule is always the same, i.e., pick the chain that contains the most aggregate work. From the perspective of the transaction fee revenues, in Section~\ref{sec::transaction_fee_split_rule} and Appendix~\ref{app::fee_split_attack}, we show that such an attack is not profitable for the adversary.

\section{Analysis of the Double-Spending Attack}\label{app::double_spend}
\subsection{The Attack} 
Essentially, we have to analyze the following. At block $h+1$, the miners start mining (voting) on ledgers that are created by miners of the proofs at block $h$. When an honest ledger of block $h$ receives $L$ votes at block $h+1$ and is part of the heaviest chain, the ledger is confirmed. See for example $\omega_6^h$ in \figref{fig::our_parallel_protocol}, which is confirmed when block $h+1$ is fully created. Any time after the honest ledger is confirmed, an adversarial ledger from block $h$, for example $\omega_7^h$ in \figref{fig::our_parallel_protocol}, is confirmed and replaces the honest ledger, if a heavier chain is published. Such a replacement is an instance of double-spending in our analysis.

We first recall that a double-spending analysis is more complicated than simply analyzing a private attack when network delays are present. In other words, when network delays are present, the views of honest miners differ and some honest miners can be lured into working for the adversary temporarily during delays, increasing its hashrate, known as balanced attacks \cite[Lemma~F.~1]{nakamoto-always-wins}. This in turn complicates the analysis. 

Without network delays, a private attack can be shown to succeed if any other strategy succeeds in Nakamoto consensus \cite[Theorem~5.1]{nakamoto-always-wins}. This result is easily extendable to the heaviest chain rule here if we had no network delays. To analyze double-spending under network delays, we will give the adversary certain advantages that simplify our analysis and give an upper bound for the double-spending probability. In other words, we will provide achievable consistency results from the honest miners' perspective. 

Here, transaction fee fork choice rule does not matter as the most aggregate rule is always in place and we break ties in the favor of the adversary. $\Delta$ considered here can be seen as $2$ times the original $\Delta$ value in the protocol description since the adversary can avoid releasing its ledger if it offers the most fee resulting in doubling of a valid ledger propagation time. We ignore such a triviality here as it does not effect our analysis.

\subsection{Analysis} 
Assume at time $\tau$, the block $h$ has $L$ proofs and a ledger of these proofs are starting to be disseminated through the network. At this point, the honest miners need $\Delta$ time to receive an honest ledger of block $h$. From this point onward, honest miners will have to create a complete block at height $h+1$ with $L$ (or $L+1$) proofs each of which will experience a $\delta$ delay. When these proofs are mined and dispersed, another $\Delta$ time is needed for the proposed ledger of block $h+1$ to be received by honest miners. We call the time the ledger of block $h+1$ is received by all honest miners as $\tau_c$. Notice that, at this point, the ledger of the block $h$ is confirmed according to $1$-block confirmation rule since block $h+1$ is complete. Thus, the interval $[\tau,\tau_c]$ is called confirmation interval.

First, note that at time $\tau$, the adversary might have a private chain that has more work than the public work observable by the honest miners. This initial adversarial advantage is called the lead. We call the interval where the adversary mines to try to build a lead as the pre-mining phase as in \cite{guo-btc-sec-lat,our-sec-lat-extended}. At $\tau_c$, when the ledger is confirmed, if the adversary has a private chain with more work, it can release it to convince honest miners to switch to the adversarial ledger, which will undo the honest work. Else, it can still keep mining on its own chain in the hopes of catching the honest work and replacing their ledger. This interval after $\tau_c$ is called post-confirmation race as in \cite{guo-btc-sec-lat,our-sec-lat-extended}.

We start by assuming that every honest proof experiences maximal $\delta$ delay (and the ledger experiences maximal $\Delta$ delay) and any honest proof generated within that delay is converted to an adversarial proof, where the adversary can decide where to put it. In other words, when an honest proof is generated, within the following $\delta$ time, honest miners are assumed to stop mining and the adversarial hashpower rises from $\alpha\cdot HP$ to $1\cdot HP$ for this $\delta$-time, where we denote full hashpower of the system as $HP$. This additional adversarial advantage is inspired by the rigged model of \cite{our-sec-lat-extended} which is a refined version of the rigged model considered in \cite{guo-btc-sec-lat}.

Notice that, when we assume the rigged model in our protocol, there will not be any honest parallel proofs in the system as the parallel proofs are only generated due to the network delay honest miners experience. As a result, we need $L$ sequential proofs for the block confirmation, which is no different than $L$-block confirmation rule in Nakamoto consensus. For example, in Fig.~\ref{fig::our_parallel_protocol}, the ledger $\omega_8^{h+1}$ is confirmed when block $h+2$ is created which has $L$ sequential proofs. This is no different than asking $L$ blocks to be mined on top of the block to be confirmed in Bitcoin. As the honest miners work for the adversarial favor during the $\delta$ and $\Delta$ delays, all honest proofs will be on different heights (AHPODH, see AHBODH in \cite{our-sec-lat-extended,guo-btc-sec-lat}). 

\begin{lemma}
    {\normalfont \textbf{(Guo-Ren\cite{guo-btc-sec-lat})}} Under AHPODH, if any attack succeeds in violating a transaction's safety, then the private mining attack also succeeds in violating that transaction's safety.
\end{lemma}

We skip proving this result as it essentially follows the same steps as bounded $\Delta$-delay model, by simply replacing blocks with proofs and $\Delta$ with $\delta$ in \cite[Lemma~5]{guo-btc-sec-lat}. We note that, since our protocol allows two parallel proofs to be combined, the balanced attacks are less likely to succeed and the rigged model assumption we make here is an extremely pessimistic one. Further, \cite{Impact_of_Temporary_Fork,Impact_of_Network_Connectivity_on_Consensus,sakurai2024modelbasedanalysisminingfairness} show that more than two proofs being mined within a $\delta$ time is extremely unlikely, hence adversarial balanced attacks are less effective in our protocol.

Let $\lambda_B=\frac{1}{T_B}$ be the block generation rate where $T_B$ denotes the block interarrival time. Let $\lambda_P$ denote the proof generation rate. Let $\beta$ denote the fraction of honest hashpower in the system. As honest miners need $L$ sequential proofs in the rigged model, we assume 
\begin{align}
    \lambda_P=L\lambda_B,
\end{align} and 
\begin{align}
    p=1-q=\beta e^{-\lambda_P\delta}>\frac{1}{2}.
\end{align}
Let $S_{L}\sim NegBin(L,\beta)$ denote the sum of $L$ i.i.d. geometric random variables with success probability $\beta$. Let $S_{\delta_L}\sim\sum_{i=1}^{L}Poisson(\lambda_P\delta)$ denote the sum of $L$ i.i.d.~Poisson random variables with rate $\lambda_P\delta$, whereas $S_{\Delta_2}\sim\sum_{i=1}^{2}Poisson(\lambda_P\Delta)$. 

\begin{lemma}
    Under the private attack with the rigged model, the adversarial proofs generated in the confirmation interval are bounded by the distribution of
    \begin{align}
        S_{L}+S_{\delta_L}+S_{\Delta_2}.
    \end{align}
\end{lemma}

\begin{Proof}
    At $\tau$, $\Delta$ time is needed for the proposed honest ledger of block $h$ to be received by the honest miners, during which they work for the adversary. Hence, the total number of proofs created during the delay is $Poisson(\lambda_P\Delta)$. Next, before the arrival of each honest proof, the adversary creates $Geo(\beta)$ proofs where $Geo(\beta)$ is a geometric random variable with success probability $\beta$. After the arrival, the honest proof suffers $\delta$ delay during which $Poisson(\lambda_P\delta)$ proofs are generated, all of which belonging to the adversary since honest miners work in its favor. Once $L$ such proofs are generated, a proposed ledger for block $h+1$ will be shared in $\Delta$ time for the block to be considered complete, during which honest miners work for the adversary.
\end{Proof}

Note that before $\tau$, the adversary might also build some initial lead for the attack \cite{sompolinsky2016bitcoin}, which we call pre-mining attack. The analysis of such a lead normally does not require a rigged model since balanced attacks are only possible after $\tau$ \cite{cao2023tradeoff}, however, we simply take the rigged model of \cite{guo-btc-sec-lat} giving another additional power to the adversary here as the lead is insignificant. We also assume that the starting point of the adversarial lead is somewhere in the proof tree of block $h$ and the block tree of block $h$ is of infinite length. 

This mild assumption essentially allows us to ignore the ledger delay of the block $h-1$ and incorporate all proofs before $\tau$ into a single proof tree. In any case, it seems nearly impossible for the adversary to sustain a lead for more than $L$ proofs such that the starting point of this lead is more than $1$ block away. We could consider this additional detail by considering a more general ruin theoretical analysis introduced in \cite{our-random-delay}, however it would merely change the insignificant digits in the final result. We refer the reader to \cite{guo-btc-sec-lat,our-random-delay} and simply note the following result.
\begin{lemma}
    The adversarial lead at $\tau$ can be upper bounded as 
    \begin{align}
        M_{lead}\sim Geo\Big(1-\frac{q}{p}\Big).
    \end{align}
\end{lemma}

\begin{Proof}
We refer the reader to \cite[Section~5.1]{guo-btc-sec-lat} for the details of the proof. Essentially, we can assume $\tau$ to be large and the lead becomes the steady-state distribution birth-death process since it cannot go below zero, where the probability of a step forward is $q=1-p$, i.e., an adversarial arrival. 
\end{Proof}

At $\tau_c$, when the honest ledger is confirmed, the adversary has $M_{lead}+S_{L}+S_{\delta_L}+S_{\Delta_2}$ proofs. If this is greater than $L$ proofs the honest miners have for the confirmation of the honest ledger, then the adversary can simply release the private proofs and undo the honest ledger, i.e., double-spend. Else, it still has a chance to catch up, in the post-confirmation race. 

Again, for simplicity, we assume the same rigged model of \cite{guo-btc-sec-lat}, even though it is not possible for the adversary to mount a balanced attack after $\tau_c$ anymore \cite{cao2023tradeoff}. Making the same mild assumption of ignoring the ledger delays after block $h+2$ for simplicity and incorporating all the proofs to be mined after $\tau_c$ into a single proof tree, the number of proofs that the adversary can make up for its deficit is denoted as $M_{deficit}\sim M_{lead}$; see \cite{guo-btc-sec-lat,our-sec-lat-extended,our-random-delay,cao2023tradeoff}.

Thus, ignoring the $\Delta$ ledger delays before block $h-1$ and after $h+2$ (we still consider the $\delta$ delays of all proofs for all heights), we have the following theorem \cite{our-random-delay}.

\begin{theorem}
    Given proof mining rate $\lambda_P$, honest fraction $\beta$, ledger delays of $\Delta$, proof delays of $\delta$ and $1$-block confirmation rule, the confirmed ledger of block $h$ with $L$ votes cannot be discarded with probability greater than,
    \begin{align}
    P(M_{lead}+S_{L}+S_{\delta_L}+S_{\Delta_2}+M_{deficit}\geq L). \label{safety-vio}
    \end{align}
\end{theorem}

\section{MDP Models for the Proposed Protocol}\label{app::mdp}
In this section, we assume the reader is familiar with the $\epsilon$-optimal selfish mining strategies of \cite{optimal-selfish}. We avoid presenting the same average reward ratio MDP model and the results. We simply refer the reader to the relevant parts of the paper \cite[Section~2, Section~4.1]{optimal-selfish}.

To consider proof withholding attacks in our protocol, first note that there are two types of proofs, initiator and incremental proofs. We need to take the different proof types into account when considering an MDP model. Assume a last common proof (or two proofs if parallel) present in both adversarial and honest chain. Right after this proof, when the adversary creates a private branch, the analysis depends on the first proof height that was withheld, until the adversary releases and overrides the honest branch, or adopts the honest branch.   

Let us assume the withholding attack, i.e., adversarial private branch, starts on an incremental proof. This means the previous proof height is a common proof between adversarial and honest branches. In this scenario, if the adversary releases the incremental proofs after honest miners mine: 
\begin{enumerate}
    \item Only $1$ incremental proof: The honest miners can combine the first adversarial and honest proofs as they are on the same proof height and share the same ancestor, and rewards will be split. In such a scenario, the honest branch length essentially increases by one, but also contains a fraction of adversarial rewards. If the adversarial branch had $2$ proofs, network split happens. If it had more, it can override.
    \item More than $1$ proof: The adversary can override the honest branch if it has more than the honest proofs in its private branch. If it has the same number of proofs, a network split of $\gamma$ happens between honest miners. In such a scenario, the parallel combination of the first adversarial and honest proof is not relevant.
\end{enumerate}

First assume $L\rightarrow \infty$, i.e., virtually all proofs are incremental proofs. Notice that, in this situation, i.e., when the proofs with same heights can always be combined together to increase aggregate chain work and rewards are split, we only need to reconfigure the MDP model of \cite{selfish-mining} to analyze the resilience against incentive attacks where we can ignore the concept of blocks and only focus on proof DAG. Hence, we modify the average reward ratio MDP of \cite{optimal-selfish} to incorporate the effect of the reward splitting, which we call reward split MDP. 

We consider the following modified objective function representing the \textit{relative payoff},
\begin{align}
    \alpha_{rel}=\frac{\sum_{t}A_t-\frac{A'_{t}}{2}}{\sum_{t}A_t+H_t-A'_{t}}, \label{eq::rel_rew_payoff_objective}
\end{align}
where $A_t$ and $H_t$ are the number of the adversarial and honest proofs that make into the heaviest chain at step $t$ and $A'_t$ is the number of proof heights that contain parallel proofs, i.e., those proofs whose coinbase rewards are split between adversarial and honest parallel proofs. As in \cite{optimal-selfish,selfish-mining,stubborn-mining,prob-selfish-mdp-method}, we consider $\gamma$ network influence of the adversary. In other words, whenever honest miners mine a proof, the adversary can match it if it has a proof with the same aggregate work and the honest network will split between the two if these two proofs cannot be combined, i.e., when they do not have the same parents. 

We consider the same action space as in \cite{optimal-selfish} but expand the state space to $(a,h,fork,p)$  with $p$ representing if parallel proofs exist. Here, we only mention the modifications to the MDP of \cite{optimal-selfish}. The parameter of parallel proofs at each step can be restricted to $p\in\{0,1\}$ since parallel proofs only happen if the adversary and honest miners work on top of the same proof/proofs of the previous proof height, which implies an \textit{adopt} happened in the last step. When $(a,h,fork)=(1,1,relevant)$, a match action will make $p=1$ since the role of $\gamma$ becomes irrelevant as both proofs can be combined by honest miners and $h+p$ becomes the aggregate work of the honest chain that the adversary has to race with if it does not adopt the combination of the two parallel proofs immediately. Finally, we modify the reward function of \cite{optimal-selfish} as 
\begin{align}
    \omega_\rho(x,y,p)=(1-\rho)x-\rho y+(-0.5+\rho)p,
\end{align} 
according to \eqref{eq::rel_rew_payoff_objective} where $x$ and $y$ are the number of adversarial and honest proofs that make into the heaviest chain at the end of each step and $p\in\{0,1\}$ represents if the reward of the first proof is split between parallel proofs.

Next, when $L$ is finite, to consider the effect of the initiator proofs, assume the withholding attack, i.e., adversarial private branch, starts on an initiator proof. This means that the previous proof height is the last proof height of the previous block. This in turn implies that, even though the previous proofs of the previous proof height is a common ancestor for both the adversarial and honest branches, the initiators can only be combined if they vote on the same ledger for the previous block. Then, if:
\begin{enumerate}
    \item The honest and adversarial initiators vote on the same valid ledger (does not matter if the ledger is created by the adversary as long as it is valid), the situation is no different than the previous analysis with incremental heights.
    \item If the adversary and honest miners vote on different ledgers, the adversarial initiator cannot be combined with the honest initiator. Thus, until an override or adopt action happens, no parallel proofs can be combined, and the effect of $p$ introduced in our model is suspended until the next private branch. In other words, the system would temporarily have the same rules as Nakamoto consensus and the MDP model of \cite{optimal-selfish} could be used directly. We can expand the state space to $(a,h,fork,p,l)$ with $l$ keeping track of the last common proof height in order to check if this situation happens.  
\end{enumerate}

To consider if this last case of expanded state space helps the adversary or not, we can first compare the relative rewards of MDP model of \cite{optimal-selfish} and the reward split MDP we explained earlier. If the reward split MDP results in higher relative rewards for an $(\alpha,\gamma)$ pair, it makes no sense for the adversary to withhold the ledger with the most fees. On the other hand, if the reward split MDP results in less relative rewards, then it is in the interest of the adversary to launch this type of attack for initiator heights if it has a ledger offering the most fees. 

We assume the adversary can always create a ledger with most fees offered, which is more than what the adversary could achieve. We implement this complicated situation where we have a state space of the form $(a,h,fork,p,l)$ with $l$ keeping track of the last common proof height. Even in the scenario where there is no reward loss, the result is approximately a weighted average between the reward ratio of the work in \cite{optimal-selfish} (with approximate weight $1/L$) and reward split MDP (with approximate weight $(L-1)/L$). This is expected since the last case arises in approximately $\frac{1}{L}$ of the proof heights, i.e., the initiator heights. On the other hand, for $\gamma<0.3$, the result is the same as the reward ratio of reward split MDP since it is more profitable to have parallel proofs in this case. As the gain is minor even when the adversarial losses are ignored, we omit the simulation with expanded state space as the computational burden becomes very large with increasing $L$. In any case, initiator proof heights are only $\frac{1}{L}$ of the total proof heights in the system.

\begin{figure}[ht!]
     \centering
     \begin{subfigure}[b]{0.48\columnwidth}
         \centering    \includegraphics[width=\textwidth]{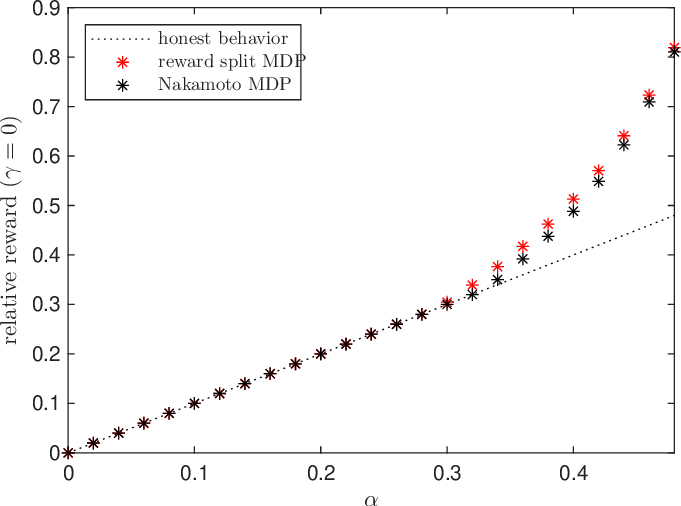}
         \caption{$\gamma=0$}
        \label{fig::mdp_comp_gamma_0}
     \end{subfigure}~
     \begin{subfigure}[b]{0.48\columnwidth}
         \centering    \includegraphics[width=\textwidth]{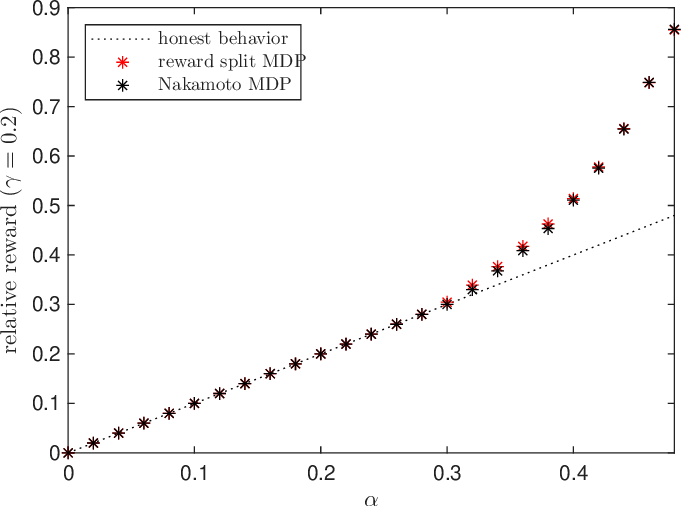}
         \caption{$\gamma=0.2$}
        \label{fig::mdp_comp_gamma_2}
     \end{subfigure}
     \hfill
     \begin{subfigure}[b]{0.48\columnwidth}
         \centering    \includegraphics[width=\textwidth]{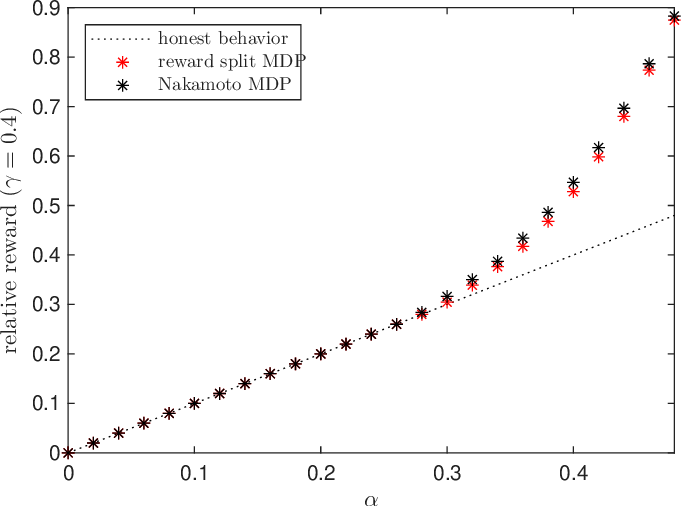}
         \caption{$\gamma=0.4$}
        \label{fig::mdp_comp_gamma_4}
     \end{subfigure}~
     \begin{subfigure}[b]{0.48\columnwidth}
         \centering    \includegraphics[width=\textwidth]{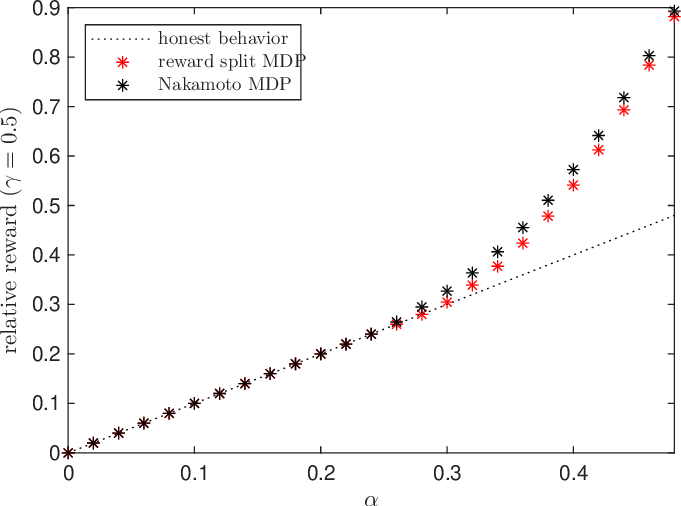}
         \caption{$\gamma=0.5$}
        \label{fig::mdp_comp_gamma_5}
     \end{subfigure}
     \hfill
     \begin{subfigure}[b]{0.48\columnwidth}
         \centering    \includegraphics[width=\textwidth]{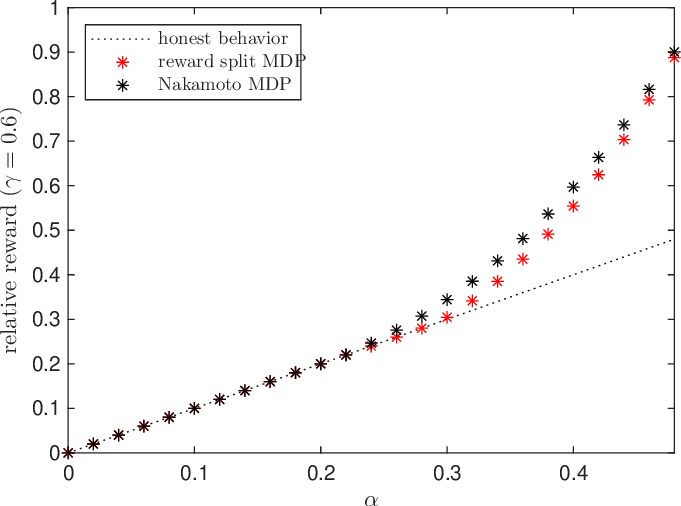}
         \caption{$\gamma=0.6$}
        \label{fig::mdp_comp_gamma_6}
     \end{subfigure}~
     \begin{subfigure}[b]{0.48\columnwidth}
         \centering    \includegraphics[width=\textwidth]{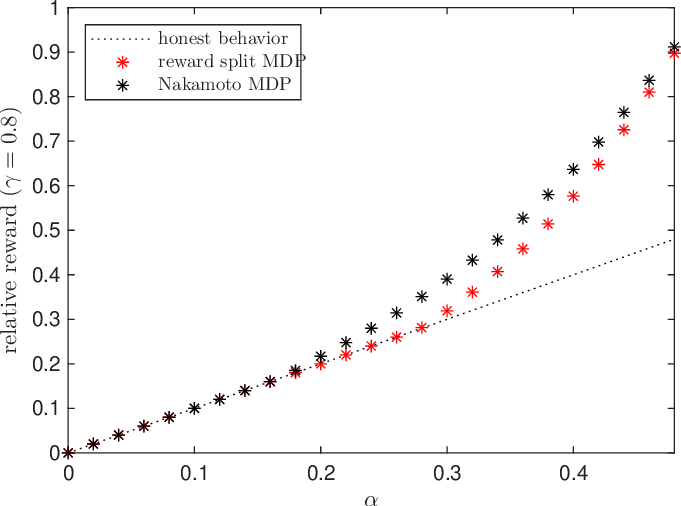}
         \caption{$\gamma=0.8$}
        \label{fig::mdp_comp_gamma_8}
     \end{subfigure}
     \caption{Reward split MDP vs Nakamoto MDP \cite{optimal-selfish}.}
	 \label{fig::simulations_mdps}
\end{figure}

\subsection{Simulations}
To evaluate the performance of the reward split MDP and compare it with the Nakamoto consensus MDP of \cite{optimal-selfish}, we run the MDP models for $\gamma=0,0.2,0.4,0.5,0.6,0.8$ and $\alpha$ values range from $0.02$ to $0.48$ with $0.02$ spacing between them with $10^{-5}$ precision. From \figref{fig::simulations_mdps}, it is clear that, reward splitting helps adversary when $\gamma$ is low as matching always gives a half reward when $(a,h,fork)=(1,1,relevant)$, whereas the adversarial relative reward in Nakamoto consensus increases with increasing $\gamma$ since matching succeeds more often. Thus, even when $\gamma=0.8$, the profitability threshold in our protocol stays around $\alpha\approx0.28$.

\section{Transaction Split Rule}\label{app::fee_split_attack}
Here, we consider an attacker that wishes to increase its rewards from the transactions fees by picking its own proposed ledger in the previous block instead of the ledger offering the most fees. We show the conditions for the attack to be profitable. Further, we show that such an attack does not have a big impact on the transaction fee rewards received by other miners.

More concretely, assume a miner $M$ holds $\alpha$ fraction of the hashpower in the system. Further, assume that $L$ proofs are created for the last block height $h$ and the miner $M$ was lucky enough to create at least one proof for the block height $h$ (hence, a proposed ledger for next block). Let $F_M$ denote the maximum offered transaction fee by the miner $M$'s proposed ledger, whereas $F_{M'}$ denotes the maximum offered transaction fee overall by any proposed ledger for the next block height. Ignoring the small $\delta$ delays of the proofs, we have the following result.

\begin{lemma}
    It is in the interest of miner $M$ to create an initiator proof referring to its own proposed ledger instead of the ledger offering the most fees if
    \begin{align}
        \frac{F_M}{F_{M'}}\geq 1-\frac{r}{r+\frac{1-r}{L}(1+(L-1)\alpha)}. \label{eq::tx_fee_comparison}
    \end{align}
\end{lemma}

\begin{Proof}
    Assume the miner picks its own proposed ledger while trying to mine an initiator proof for the next block height:
    \begin{enumerate}
        \item If it is successful to mine an initiator referring to the ledger offering $F_{M}$ fees before others, which happens with probability $\alpha$, it shares the initiator immediately. Other miners accept the initiator of $M$ and mine on top of it since it is the chain with the heaviest work.
        \item If others mine an initiator before him referring to the ledger offering $F_{M'}$ fees, which happens with probability $1-\alpha$, the miner $M$ accepts the ledger and mines on top of it as everyone else.
    \end{enumerate} 
    
    For the rest of the $L-1$ incremental proofs after the initiator, the miner $M$ creates $(L-1)\alpha$ more proofs on average. The expected transaction fees rewards is 
    \begin{align}
        &F_M\cdot\alpha\left(r+\frac{1-r}{L}+\frac{1-r}{L}(L-1)\alpha\right)+F_{M'}\cdot(1-\alpha)\left(\frac{1-r}{L}(L-1)\alpha\right).
    \end{align}
    
    On the other hand if it simply tries to mine an initiator proof referring to the ledger offering $F_{M'}$ fees, the expected transaction fees rewards is
    \begin{align}
        F_{M'}(1-r)\cdot\alpha.
    \end{align}
    Comparing the two strategies, we get \eqref{eq::tx_fee_comparison}.
\end{Proof}

Notice that, \eqref{eq::tx_fee_comparison} is increasing in $\alpha$. Hence, when the attacker has more fraction of hashpower, its own ledger has to offer similar fees as the ledger offering the most fee. For example, even if we picked an equal split rule between the proposer and voters, i.e., $r=\frac{1}{{C(\omega_L^h)}+1}$, \eqref{eq::tx_fee_comparison} becomes
\begin{align}
    \frac{F_M}{F_{M'}}\geq \frac{\left(L-1\right)\alpha+1}{\left(L-1\right)\alpha+2}.\label{eq::tx_fee_eq_split}
\end{align}

If $\alpha=0.1$ and $L=50$, the threshold in \eqref{eq::tx_fee_eq_split} becomes $0.85$, i.e., the ledger of $M$ should offer at least $0.85F_{M'}$, else the attack is not profitable. On the other hand, with equal split rule as $\alpha\rightarrow 0$, if $M$ is lucky enough to propose a ledger, it is in its own interest to mine an initiator referring to its own ledger as long as it offers at least half the rewards compared to the ledger offering the most fees. In other words, the ledger of $M$ should offer at least $0.5F_{M'}$. Hence, we do not pick equal split rule.

If we pick smaller values for $r$ compared to equal split rule, for example pick $r=\frac{1}{2L}$, for $L=50$, if the ledger of $M$ offers at least $0.85F_{M'}$, it needs to hold less than $\alpha\leq0.038$ fraction of hashpower in the system. Hence, as we decrease $r$, the attack becomes unprofitable even for a small miner. On the other hand, even when it is profitable, as long as $r$ is small, the transaction fee rewards for the other miners that create incrementals on top of this ledger do not lose much, for example, with $r=\frac{1}{2L}$ and $\alpha=0.1$ they only lose $0.08F_{M'}\frac{1-r}{L}$. 

We also note that the above analysis assumes that the attacker was lucky enough to mine a proof with a proposed ledger for the previous block. Even though the threshold in \eqref{eq::tx_fee_comparison} decreases with $\alpha$, the frequency of such a lucky situation also decreases with $\alpha$. 

\bibliographystyle{ieeetr}
\bibliography{blockchain}

\end{document}